\documentclass[proof]{WileyASNA-v1}
\usepackage{framed}
\usepackage{natbib}

\usepackage{romannum}

\usepackage{soul}

\AtBeginDocument{\pagenumbering{arabic}}

\articletype{Original Article}

\received{Date}
\revised{Date}
\accepted{Date}

\begin{document}

\title{Modeling brightness temperature of prominences on the solar disk using ALMA single-dish observations}

\author[1]{Filip Matkovi\'c*}

\author[1,2]{Roman Braj\v{s}a}

\author[3,4]{Arnold O. Benz}

\author[2]{Hans -G. Ludwig}

\author[5,6]{Caius L. Selhorst}

\author[7,8]{Ivica Skoki\'{c}}

\author[1]{Davor Sudar}

\author[9]{Arnold Hanslmeier}

\authormark{Matkovi\'{c} \textsc{et al}}

\address[1]{\orgdiv{Hvar Observatory}, \orgname{Faculty of Geodesy, University of Zagreb}, \orgaddress{\state{Zagreb}, \country{Croatia}}}

\address[2]{\orgdiv{Landessternwarte}, \orgname{Zentrum f\"ur Astronomie der Universit\"at Heidelberg}, \orgaddress{\state{Heidelberg}, \country{Germany}}}

\address[3]{\orgdiv{University of Applied Sciences and Arts Northwestern Switzerland}, \orgaddress{\state{Windisch}, \country{Switzerland}}}

\address[4]{\orgdiv{Institute for Particle Physics and Astrophysics}, \orgaddress{\state{ETH Zurich}, \country{Switzerland}}}

\address[5]{\orgdiv{NAT - N\'ucleo de Astrof\'isica}, \orgname{Universidade Cidade de S\~ao Paulo}, \orgaddress{\state{S{\~a}o Paulo}, \country{Brazil}}}

\address[6]{\orgdiv{Center for Solar-Terrestrial Research}, \orgname{New Jersey Institute of Technology, Newark}, \orgaddress{\state{New Jersey}, \country{USA}}}

\address[7]{\orgdiv{Astronomical Society} \orgname{"Anonymus"}, \orgaddress{\state{Valpovo}, \country{Croatia}}}

\address[8]{\orgdiv{Protostar Labs} \orgname{d.o.o.}, \orgaddress{\state{Beli{\v s}\'ce}, \country{Croatia}}}

\address[9]{\orgdiv{Institute of Physics}, \orgname{University of Graz}, \orgaddress{\state{Graz}, \country{Austria}}}

\corres{*Filip Matkovi\'{c}, Hvar Observatory, Faculty of Geodesy, University of Zagreb, Ka\v{c}i\'{c}eva 26, 10000 Zagreb, Croatia. \\\email{fmatkovic@geof.hr}}

\presentaddress{Croatian Science Foundation, project ID: 7549. Austrian-Croatian Bilateral Scientific Projects. Horizon 2020 project SOLARNET, project ID: 824135. Alexander von Humboldt Foundation. S{\~a}o Paulo Research Foundation (FAPESP), grant number: 2019/03301-8.}

\fundingInfo{Croatian Science Foundation, project ID: 7549. Austrian-Croatian Bilateral Scientific Projects. Horizon 2020 project SOLARNET, project ID: 824135. Alexander von Humboldt Foundation. S{\~a}o Paulo Research Foundation (FAPESP), grant number: 2019/03301-8.}

\jnlcitation{\cname{%
\author{Matkovi\'c F.}, 
\author{Braj{\v s}a R.}, 
\author{Benz A. O.},
\author{Ludwig H.-G.},
\author{Selhorst C. L.},
\author{Skoki\'c I.},
\author{Sudar D.}, and 
\author{Hanslmeier A.}} (\cyear{Year}), 
\ctitle{Modeling brightness temperature of prominences on the solar disk using ALMA single-dish observations}, \cjournal{Astronomische Nachrichten}, \cvol{Vol. No.}.}

\abstract{Prominences (PRs) are among the most common solar phenomena, yet their full physical picture, particularly their chromospheric mm emission, remains incomplete. The new Atacama Large Millimeter/submillimeter Array (ALMA) presents an opportunity to study PRs at mm and sub-mm wavelengths through a combination of measurements and theoretical modeling. We utilize ALMA single-dish measurements alongside data from other radio instruments to model the PR brightness temperature through adaptation and modification the 1D semi-empirical Avrett-Tian-Landi-Curdt-W\"ulser (ATLCW) quiet-Sun (QS) model. The calculated and measured PR brightness temperatures were found to be lower than the measured QS value and predictions from the unperturbed ATLCW QS model across the ALMA wavelength range, consistent with PRs appearing in absorption. The PR density was found to be 60 -- 163 times higher and temperature 155 -- 163 times lower than the QS level, aligning with previous measurements. A key finding emerged with the non-hydrostatic equilibrium assumption, yielding a more physically consistent PR brightness temperature. This suggests that PR stability is most likely maintained by its magnetic field obeying magnetostatic conditions rather than by pure hydrostatic equilibrium, supporting recent studies. Additionally, our results confirm that thermal bremsstrahlung is the dominant radiation mechanism for PRs at mm and sub-mm wavelengths.}


\maketitle

\section{Introduction}
\label{Introduction}
Prominences (PRs) are one of the most common features in the solar atmosphere. When observed above the solar limb, they appear as bright arched loops or thread-like structures, while on the solar disk, they take the form of dark, elongated clouds with a central spine and lateral extensions, commonly referred to as filaments (e.g., Ru{\v z}djak \& Tandberg-Hansen \citeyear{Ruzdjak1990}; Tandberg-Hansen \citeyear{Tandberg1995}; Parenti \citeyear{Parenti2014}). For simplicity, hereafter we use the term PR to refer to both PRs and filaments, as they represent the same solar structure viewed against different background. PRs can extend into the lower corona, where they are visible in the optical spectrum and cooler extreme-ultraviolet (EUV) spectral lines. Composed of cooler, denser chromospheric plasma, PRs are typically observed in absorption, making them thermally and pressure-isolated from their surroundings (e.g., Ru{\v z}djak \& Tandberg-Hansen \citeyear{Ruzdjak1990}; Tandberg-Hansen \citeyear{Tandberg1995}; Parenti \citeyear{Parenti2014}; Brajša et al. \citeyear{Brajsa2018a}).

Based on the most common EUV and H$\alpha$ observations, the typical plasma temperature in PRs ranges from 7~500 to 9~000 K, while the PR electron density typically falls between $10^9$ and $10^{11}$ cm$^{-3}$ (Parenti \citeyear{Parenti2014} and references therein), with rare cases reaching up to $10^{15}$ cm$^{-3}$ (Lites et al. \citeyear{Lites2010}). The cool and dense plasma of PRs is embedded in a magnetic environment located above the photospheric magnetic inversion lines and, in their quiescent state, PRs maintain equilibrium, allowing their mass to be supported within the tenuous corona (e.g., Parenti \citeyear{Parenti2014}). It is assumed that the magnetic field that permeates the PR structure provides the necessary framework for plasma equilibrium, ensuring PR stability despite external forces (e.g., Gibson \citeyear{Gibson2018} and references therein). However, due to limited observational data, the exact mechanisms and magnetic configurations responsible for their stability remain unknown.

Theoretical models play a crucial role in improving our understanding of PR properties and their observed appearance. Various approaches exist for modeling solar structures, but they generally fall into two main categories: purely theoretical or partially observation-based 3D numerical modeling and observation-based semi-empirical modeling. Since PRs are inherently 3D structures extending into coronal heights, they are typically modeled using the first approach. This involves purely theoretical magnetohydrodynamic simulations and radiation emission and magnetic field measurements to replicate the observed structure (e.g., Gibson \citeyear{Gibson2018} and references therein). However, while 3D numerical models are valuable for studying the PR topology and visual appearance, they cannot accurately reproduce the physical values obtained from measurements. Instead, they are primarily used to simulate structural characteristics observed in actual data.

In this work, we aim to derive physical values for the PR properties that are consistent with measurements. To achieve this, we use the second approach based on the semi-empirical atmospheric models. These models are typically constructed from observed spectral lines of various elements, primarily in the EUV and infrared parts of the electromagnetic spectrum. For radio emission, semi-empirical models have been mainly used to describe solar background structures like the quiet Sun (QS) and, in modified forms, active regions (ARs) and coronal holes (CH). Notable models like the Vernazza-Avrett-Loeser (VAL) (Vernazza et al. \citeyear{Vernazza1981}), Fontenla-Avrett-Loeser (FAL) (Fontenla et al. \citeyear{Fontenla1993}), or Selhorst-Silva-Costa (SSC) (Selhorst et al. \citeyear{selhorst2005}) models have been applied to describe the brightness temperature and other plasma properties of such solar structures (e.g., Benz et al. \citeyear{Benz1997}; Braj{\v s}a et al. \citeyear{Brajsa2007}, \citeyear{Brajsa2009}; Selhorst et al. \citeyear{selhorst2017}, \citeyear{Selhorst2019}; Braj{\v s}a et al. \citeyear{Brajsa2018a}).

There have been previous PR semi-empirical modeling attempts like, for example, Braj{\v s}a et al. (\citeyear{Brajsa2009}), who utilized full-disk solar observations at 8 mm wavelength obtained with the 14-m radio telescope of the Mets\"ahovi Radio Observatory of the Helsinki University of Technology. Assuming hydrostatic equilibrium in PRs and a thermal bremsstrahlung as the dominant radiation mechanism at mm and sub-mm wavelengths, the authors applied the FAL model, modifying its density and temperature parameters, to determine the PR properties. Braj{\v s}a et al. (\citeyear{Brajsa2009}) derived PR models showing a temperature 155 -- 175 times lower and a density higher by the same factor than that of the QS, with the brightness temperature decreasing from 7\;700 K to 6\;780 K with increasing (decreasing) density (temperature) compared to the constant QS brightness temperature of about 7\;840 K. The authors also modeled the case of coronal condensation with a 175 times lower temperature and a 50 times higher density than the QS, with a brightness temperature of about 6\;880 K. Although coronal condensations can cause absorption in the mm emission, their density is not high enough to be visible in H$\alpha$ absorption (Kundu et al. \citeyear{Kundu1978}). Since the PR structure in Braj{\v s}a et al. (\citeyear{Brajsa2009}) was observed in both mm and H$\alpha$ absorption, the possibility of it being a coronal condensation is ruled out.

Due to the lack of mm and sub-mm radio PR observations, previous modeling efforts have been very limited in their ability to understand PR radio emission. Thanks to the Atacama Large Millimeter/submillimeter Array (ALMA), we now have the capability to study PRs in the wavelength range from 0.3 to 10 mm. Braj{\v s}a et al. (\citeyear{Brajsa2018a}) used ALMA observations and further developed the FAL model from Braj{\v s}a et al. (\citeyear{Brajsa2009}) to analyze the PR properties by incorporating ALMA full-disk single-dish measurements at 1.21 mm wavelength. Six PR models were created, each with progressively lower temperature and higher density than the QS by factors of 80, 120, 160, 200, 240, and 280, along with a coronal condensation model featuring 175 times lower temperature and 50 times higher density than the QS. These models, along with the corresponding brightness temperature models, were compared to the QS model and the observed PRs. As the density increased and the temperature decreased, the PR models transitioned from full emission- to full absorption-visible PRs for the largest factors. The authors also obtained a PR brightness temperature only slightly higher than the QS up to 5 mm, but well below the QS level at longer wavelengths for the coronal condensation model, which was again ruled out due to the structure being visible in the H$\alpha$ absorption.

In our most recent paper, Matković et al. (\citeyear{Matkovic2024}) (hereafter Paper \Romannum{1}), we modeled the brightness temperature in the mm and sub-mm emission for three solar background structures: QS, AR, and CH regions. The Avrett-Tian-Landi-Curdt-W\"ulser (ATLCW) atmospheric model for the QS, which served as the basis for modeling these three solar structures, provided a sufficient modeling basis and yielded results that agreed very well with both measurements and theories. Even in its original, unperturbed form, the ATLCW QS model accurately described the brightness temperature measurements, not only from ALMA but from all instruments considered for the QS region across the entire observed ALMA wavelength range. This is why this atmospheric model was initially chosen and applied. For the AR structure, the modeling based on the ATLCW QS model demonstrated that ARs are significantly brighter than the surrounding QS regions, due to the much higher density and temperature of the plasma within an AR. In contrast, for the CH structure, we found that CHs are slightly dimmer than the surrounding QS regions, primarily due to the lower plasma density, and showing insignificant dependence of the CH brightness temperature on the plasma temperature.

We now extend the modeling of the brightness temperature, plasma density, and temperature to PR structures. In Section \ref{Calculation}, we give a brief description of the methodology used to model the PR brightness temperature and other properties using data and measurements obtained with ALMA and other instruments we discuss in Section \ref{observational_results}. Finally, in Section \ref{Results}, we present a detailed description of the results from the PR modeling, which we further discuss and draw conclusions in Section \ref{discussion_conclusion}.

\section{Obtaining theoretical models and brightness temperature calculation}
\label{Calculation}

Numerous earlier studies have shown that the dominant radiation mechanism responsible for the quiescent and active chromosphere at mm and sub-mm wavelengths is thermal bremsstrahlung (Zirin \citeyear{Zirin1988}; Hurford \citeyear{Hurford1992}; Braj{\v s}a \citeyear{Brajsa1993}; White \citeyear{White2002}; Benz \citeyear{Benz2009}; Wedemeyer et al. \citeyear{Wedemeyer2016}). This assumption was also adopted in Paper \Romannum{1} for  the ALMA wavelength range. The close agreement between modeled brightness temperatures and observational data for QS, AR, and CH structures presented in Paper \Romannum{1} supports the conclusion that thermal bremsstrahlung is the primary emission mechanism in the mm and sub-mm regimes. Accordingly, here we extend this assumption to PRs, treating thermal bremsstrahlung as the main contributor to its observed mm and sub-mm emission.

In the case of thermal bremsstrahlung, the optical depth d$\tau$ at a height d$h$ and a given frequency $\nu$, neglecting the magnetic field, can be written as (e.g., Benz \citeyear{Benz2002}; Paper \Romannum{1}):
\begin{equation}
\label{Eq_optical_depth}
{\rm d}\tau_\nu=\frac{0.01146\;\mathrm{cm^{5}Hz^{2}K^{3/2}}\times g \times n^{2}_{e}}{\left(1-8.06\times 10^{7}\;\mathrm{cm^3Hz^{2}}\times n_{e}/\nu^{2}\right)^{1/2}\nu^{2}T_{e}^{3/2}} {\rm d} h,
\end{equation}
where $n_\mathrm{e}$ and $T_\mathrm{e}$ are electron density and temperature, respectively, while $g$ is the Gaunt factor (e.g., Rybicki \& Lightman \citeyear{Rybicki1985}; Weinberg \citeyear{Weinberg2020}).

In the mm and sub-mm wavelength range, the long-wavelength approximation of the Planck function applies, leading to the Rayleigh-Jeans law, which in turn states that the observed emission intensity is proportional to the temperature of the radiating body (here the solar plasma) (e.g., Wilson et al. \citeyear{Wilson2013}). If we assume this, we can write the brightness temperature $T_\mathrm{b}$ at a given wavelength $\lambda$ into a form of the radiative transfer equation (e.g., Paper \Romannum{1} and references therein):
\begin{equation}
\label{Eq_radiative_transfer}
T_\mathrm{b}(\lambda)=\int_{0}^{\infty}T_\mathrm{e}e^{-\tau_\lambda}{\rm d}\tau_\lambda,
\end{equation}
where we consider the brightness temperature as $T_\mathrm{b}(\lambda)=\lambda^4I_\lambda/(2ck_\mathrm{B})$, where $I_\lambda$ is the observed emission intensity at a given wavelength, while $c$ and $k_\mathrm{B}$ are the speed of light and the Boltzmann constant, respectively. The complete derivation of Equation \ref{Eq_radiative_transfer} is provided in Paper \Romannum{1} and references therein.

The calculation of the brightness temperature in this work is similar to the procedure described in Paper \Romannum{1}, so we only provide a brief overview of this procedure here. To calculate the brightness temperature at the observed wavelength $\lambda$, we begin with Equation \ref{Eq_radiative_transfer}, where the optical depth $\tau_\lambda$ is substituted using Equation \ref{Eq_optical_depth}, under the assumption that $c=\lambda\times\nu$. This yields a wavelength-dependent brightness temperature that depends only on the height $h$ above the solar surface, electron density $n_\mathrm{e}$ and temperature $T_\mathrm{e}$, and Gaunt factor $g$, which varies slowly with changes in electron density and temperature. In this work, the Gaunt factor is calculated using the interpolation method developed by van Hoof et al. (\citeyear{vanHoof2014}) and later implemented by Sim\~oes et al. (\citeyear{Simoes2017}) and Selhorst et al. (\citeyear{Selhorst2019}), as we did in Paper \Romannum{1}. Finally, to evaluate the brightness temperature at a given wavelength, the electron density and temperature parameters at a given height from the ATLCW QS model are inserted into Equation \ref{Eq_radiative_transfer}, integrating the contributions over all considered heights. This process yields a brightness temperature profile as a function of wavelength, which can then be directly compared with observational data. A detailed description of the full computational method is provided in Paper \Romannum{1}.

However, to obtain a brightness temperature profile that best matches the observations, we applied the widely used $\chi^2$-minimization technique (e.g., Ivezi\'c et al. \citeyear{Ivezic2014}) to find the atmospheric model that yields the best brightness temperature profile. We use the same relation for the sum of $\chi^2$ values over all wavelengths as in Equation 6 in Paper \Romannum{1}, which we want to minimize. Following the procedure in Paper \Romannum{1}, the electron density and temperature were used as variable parameters for the $\chi^2$-minimization, where for each pair of electron density and temperature values the sum of the $\chi^2$ values is calculated until the global minimum of this sum is found. The electron density and temperature parameters are taken from the original ATLCW QS model from Table 1 in Avrett et al. (\citeyear{Avrett2015}) and varied with multiplicative factors (Paper \Romannum{1}):
\begin{equation}
\label{multi_factors}
f_\mathrm{n}=\frac{n_\mathrm{e}\mathrm{(new)}}{n_\mathrm{e}\mathrm{(original)}},\; f_\mathrm{T}=\frac{T_\mathrm{e}\mathrm{(new)}}{T_\mathrm{e}\mathrm{(original)}},
\end{equation}
where $n_\mathrm{e}$ and $T_\mathrm{e}$ denote the electron densities and temperatures of the original and modified (new) ATLCW models.

Since PRs are not background structures like QS, AR, and CH in Paper \Romannum{1}, but 3D structures that extend high above the solar surface, the above multiplicative factors were applied only within a considered PR height range of $40\;000-50\;000$ km (Bastian et al. \citeyear{Bastian1993}; Braj{\v s}a et al. \citeyear{Brajsa2009}). Only within this specific atmospheric layer at the given PR height range were the electron density and temperature parameters of the ATLCW QS model varied for the purpose of the $\chi^2$-minimization. The overall integrating brightness temperature calculation procedure was then performed over all given heights in the ATLCW QS model from 0 to 57\;797 km to derive the best-fitted brightness temperature profile. Since only two height points were included within the chosen PR height range ($40\;000-50\;000$ km) in the original ATLCW QS model, inclusion of more data points through interpolation is needed to obtain a physically more accurate calculation of the brightness temperature. To increase the number of data points (i.e., dataset resolution), the height distributions of density and temperature parameters of the original ATLCW model, together with the corresponding heights, were interpolated beforehand within the entire height range (0 to 57\;797 km) of the ATLCW model using "spline" function in IDL language that performs cubic spline interpolation, where the number of data points was increased by a factor of 100. The value of 100 is chosen as the optimal interpolation factor to obtain stable (i.e., not changing anymore with increasing dataset resolution) brightness temperatures calculated throughout the ALMA wavelength range as smaller interpolation factor could result in the integration calculation of the brightness temperature greatly underestimating the brightness temperature values at long mm wavelengths.

\section{Observational results}
\label{observational_results}

\subsection{ALMA data}

In this study, we primarily utilize ALMA observations of the solar atmosphere in the mm wavelength range. Although ALMA provides single-dish (White et al. \citeyear{White2017}) and interferometric (Shimojo et al. \citeyear{Shimojo2017}) observations of the full solar disk and a smaller region on the Sun, respectively, for this work we restrict our analysis to single-dish full-disk data, which are more suitable for studying PR structures due to the larger field of view of 1\;200" in radius. Unlike the interferometric observations with a field of view limited to the beam size of a single radio antenna (e.g., at wavelengths of 1.2 and 3 mm the field of view is about 26" and 60" in diameter, respectively), single-dish observation encompasses the entire solar disk (of a radius of about 970") and, therefore, the entire PR structure with length typically in the range of roughly 80" -- 800" (Tandberg-Hanssen \citeyear{Tandberg1995}). As in Paper \Romannum{1}, we still have only two wavelength bands available, Band 3 ($\lambda\approx3$ mm) and Band 6 ($\lambda\approx1.2$ mm), while other ALMA wavelength bands (e.g., Band 5 and 7) are still in the testing phase but are expected to become available in the near future.

\begin{figure*}[h!]
\centering
\resizebox{0.99\hsize}{!}{\includegraphics{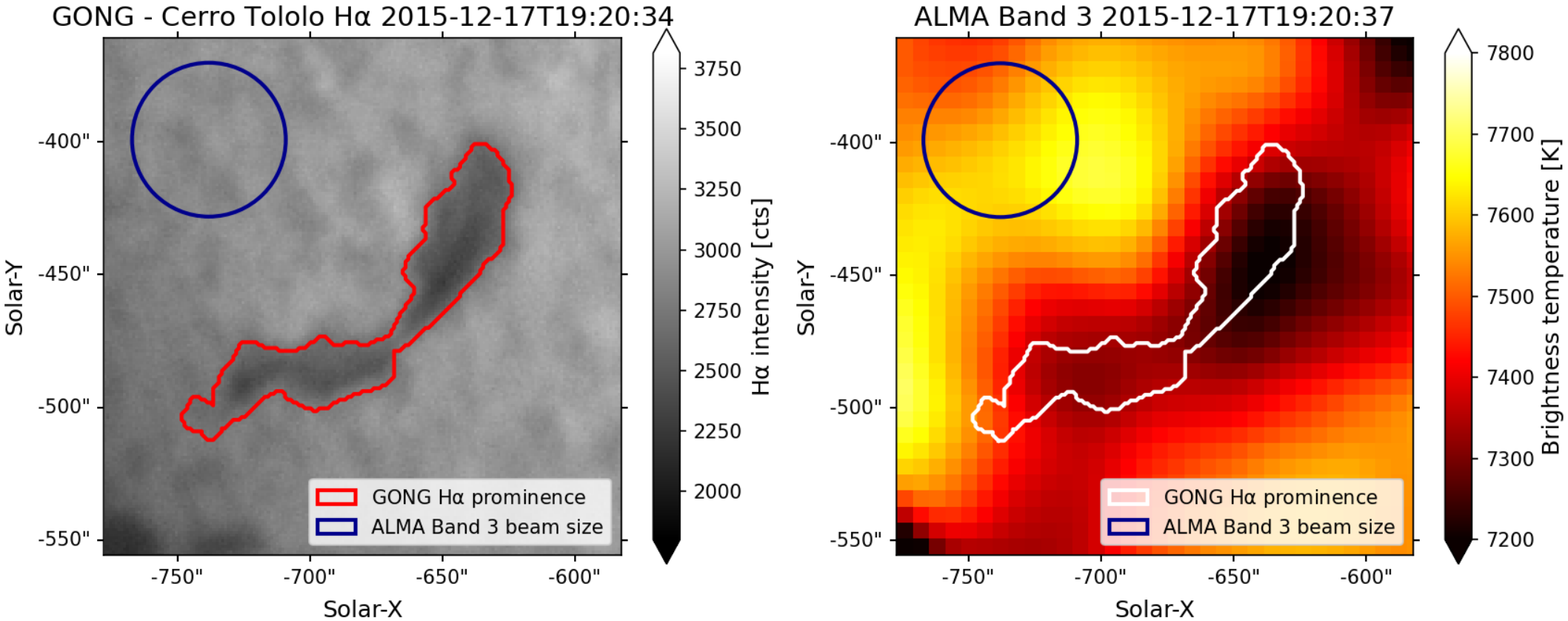}}
\caption{~Left: Cutout GONG H$\alpha$ image obtained with the Cerro Tololo optical telescope on December 17, 2015, with detected PR boundaries (red contours) and marked ALMA Band 3 beam size (blue circle). Right: Concurrent cutout ALMA Band 3 image with superimposed PR boundaries (white contours) extracted from the left H$\alpha$ image and marked ALMA Band 3 beam size (blue circle). Both images are given in helioprojective coordinate system in arcsec, where the intensity values in H$\alpha$ and ALMA Band 3 images are clipped between 1\;800 and 3\;800 cts and between 7\;200 and 7\;800 K, respectively.}
\label{PR_Halpha_ALMA_image}
\end{figure*}

To measure the PR brightness temperature, it is first necessary to determine its boundaries. Due to the relatively lower spatial resolution of ALMA single-dish observations, the contrast in emission intensity between a PR and its surroundings was poor. Therefore, to enhance this contrast and accurately identify the PR boundaries, we employ H$\alpha$ observations from the Global Oscillation Network Group (GONG)\footnote{\url{https://gong.nso.edu/}} database obtained using the optical telescope of the Cerro Tololo Interamerican Observatory. The PR boundaries were extracted from the GONG H$\alpha$ image corresponding to the time of the ALMA image and superimposed on the ALMA image as contours (Figure \ref{PR_Halpha_ALMA_image}). The PR brightness temperature in the ALMA image was then calculated by averaging the brightness temperature value across the entire area enclosed by these H$\alpha$ contours. A detailed description of the PR H$\alpha$ boundary extraction is provided in Braj{\v s}a et al. (\citeyear{Brajsa2018b}).

For the PR observed in ALMA Band 6, we adopt the brightness temperature measurement reported by Braj{\v s}a et al. (\citeyear{Brajsa2018b}). In that study, the PR (see FIL1 in Figure 1 in Braj{\v s}a et al. \citeyear{Brajsa2018b}) was observed at 1.21 mm wavelength in a full-disk image of the Sun obtained by the ALMA single-dish observation with a beam size of 26" on December 18, 2015. The corresponding brightness temperature, originally presented in Table 1 of that work, is repeated here in Table \ref{Table_data} in a modified form. In the present study, we extend the PR brightness temperature measurement by including the Band 3 measurement of the same PR structure at a wavelength of 2.80 mm obtained by the ALMA single-dish observation with a beam size of 58" on December 17, 2015 (Figure \ref{PR_Halpha_ALMA_image}), one day before the work of Braj{\v s}a et al. (\citeyear{Brajsa2018b}). The derived PR brightness temperature measurement for Band 3 is included in Table \ref{Table_data}.

When observing with ALMA, the limb brightening effect becomes apparent in the solar image and must be removed to enable accurate comparisons of emissions from different solar structures on the solar disk. There are several approaches for limb brightening correction, such as the methods of Sudar et al. (\citeyear{Sudar2019}) and Alissandrakis et al. (\citeyear{Alissandrakis2022}), who use certain fitting models of a polynomial function to the brightness temperature profile to scale down the observed emission towards the solar limb. To maintain consistency with Paper \Romannum{1}, we adopt a simple approach for the limb brightening correction, where the value of the brightness temperature of a QS region, given in Table \ref{Table_data} as $T_\mathrm{b}$(QS), is measured by averaging over a 10-pixel radius circular area located at the same radial distance from the solar center as the observed PR. This value is then used to calculate the brightness temperature difference $\Delta T_\mathrm{b}$ = $T_\mathrm{b}$(PR) $-$ $T_\mathrm{b}$(QS), where $T_\mathrm{b}$(PR) is the brightness temperature of the given PR. Simultaneously, we determine the brightness temperature $T_\mathrm{b}$(QS$_\mathrm{central}$) of a QS region of a radius of 15 pixels measured in the central part of the solar disk far from limb brightening effect. The final PR brightness temperature, corrected for limb brightening, is obtained by adding $\Delta T_\mathrm{b}$ to $T_\mathrm{b}$(QS$_\mathrm{central}$). All relevant values are presented in Table \ref{Table_data}.
\begin{table*}[h!]
\centering
\caption{~Measurements of PR brightness temperature $T_\mathrm{b}$(PR) obtained by the ALMA, Mets\"ahovi, and Nobeyama radio telescopes for a given time (Date), wavelength or frequency ($\lambda / \nu$), and spatial resolution (Beam size). The brightness temperature $T_\mathrm{b}$(QS) corresponds to the brightness temperature measurement of a QS region at a similar distance from the center of the solar disk as the PR. In the case of Mets\"ahovi and Nobeyama, $T_\mathrm{b}$(QS) $=$ $T_\mathrm{b}$(QS$_\mathrm{central}$) due to the limb brightening correction already taken into account. The brightness temperature difference $\Delta T_\mathrm{b}$, calculated as $\Delta T_\mathrm{b}=T_\mathrm{b}$(PR) $-$ $T_\mathrm{b}$(QS), is also given and used to obtain $T_\mathrm{b}^\mathrm{a}$(PR) and $T_\mathrm{b}^\mathrm{b}$(PR) measurements for procedures "a" and "b" (see Section \ref{Results} for details), given in bold in the table below, and which are plotted in Figures \ref{PR_hyd_eq} and \ref{PR_non_hyd_eq} and used for fitting the calculated brightness temperature obtained from the ATLCW QS model. Details of the individual measurements can be found in the corresponding references (Reference).}
\label{Table_data} 
\centering
\resizebox{2\columnwidth}{!}{
\begin{tabular}{c c c c c c c c c c c}
\hline\midrule
Instrument & Date & $\lambda$ / $\nu$ & Beam size & $T_\mathrm{b}$(QS) ($T_\mathrm{b}$(QS$_\mathrm{central}$)) & $T_\mathrm{b}$(PR) & $\Delta T_\mathrm{b}$ &\textbf{$\boldsymbol{T_\mathrm{b}^\mathrm{a}}$(PR)}&\textbf{$\boldsymbol{T_\mathrm{b}^\mathrm{b}}$(PR)}& Reference \\
(name) & (y:m:d) & (mm) / (GHz) & (arcsec) & (K) & (K) &  (K) &\textbf{(K)}&\textbf{(K)}& (citation)\\
\midrule
\multirow{2}{*}{ALMA}    &  \multirow{2}{*}{2015-12-18}   &  \multirow{2}{*}{1.21 / 248} & \multirow{2}{*}{26} & \multirow{2}{*}{6\;460 (6\;040)}& \multirow{2}{*}{6\;350} &  \multirow{2}{*}{$-110$}       &\multirow{2}{*}{\textbf{5\;930}}&\multirow{2}{*}{\textbf{6\;090}}&  Brajša et al. \\ 
&    &      &&  &  &  &&&(\citeyear{Brajsa2018b})    \\
\midrule
\multirow{2}{*}{ALMA}    &  \multirow{2}{*}{2015-12-17}    &  \multirow{2}{*}{2.80 / 107} & \multirow{2}{*}{58} & \multirow{2}{*}{7\;500 (7\;110)}& \multirow{2}{*}{7\;280} &  \multirow{2}{*}{$-220$}       &\multirow{2}{*}{\textbf{6\;890}}&\multirow{2}{*}{\textbf{6\;750}}&   \multirow{2}{*}{This work} \\ 
&    &      &&  &  &  &&&    \\
\midrule
\multirow{2}{*}{Nobeyama}  &  1984-07-16 (start)    &  \multirow{2}{*}{3.10 / 98} & \multirow{2}{*}{17} & \multirow{2}{*}{6\;500} & \multirow{2}{*}{6\;200} &  \multirow{2}{*}{$-300$} &\multirow{2}{*}{\textbf{$\boldsymbol{=T_\mathrm{b}}$(PR)}}&\multirow{2}{*}{\textbf{6\;750}}& Hiei et al.   \\
&  1984-07-22 (end)    &      &&  &  &  &&&(\citeyear{Hiei1986})    \\
\midrule
Mets\"ahovi  &  1993-05-27    &  8.10 / 37 & 144 & 8\;100 & 7\;890 &  $-210$         &\textbf{$\boldsymbol{=T_\mathrm{b}}$(PR)}&\textbf{7\;810}& This work   \\
\midrule
\multirow{2}{*}{Nobeyama}  &  1984-07-16 (start)    &  \multirow{2}{*}{8.30 / 36} & \multirow{2}{*}{46} & \multirow{2}{*}{8\;000} & \multirow{2}{*}{6\;600} &\multirow{2}{*}{$-1\;400$}         &\multirow{2}{*}{\textbf{$\boldsymbol{=T_\mathrm{b}}$(PR)}}&\multirow{2}{*}{\textbf{6\;660}}& Hiei et al.   \\
&  1984-07-22 (end)    &      &&  &  &  &&&  (\citeyear{Hiei1986})  \\
\midrule
\end{tabular}
}
\end{table*}

\subsection{Other radio instruments}
Since we only have two short wavelengths available for ALMA data, we are missing the long-wavelength tail from 3 to 10 mm in the ALMA range. This lack of observations at longer wavelengths poses a problem in theoretical modeling, as it may result in inaccurate interpretations or characterizations of the observed solar structures. While awaiting the availability of additional ALMA bands, we address this limitation by supplementing the existing ALMA data with data from other radio telescopes operating within the same wavelength range.

The first supplementary data come from the 14-m diameter Cassegrain-type radio telescope at the Mets\"ahovi Radio Observatory\footnote{\url{https://www.aalto.fi/en/metsahovi-radio-observatory}}, which provides full-disk and partial mapping of the Sun in the 3 mm -- 3 cm wavelength range (Urpo et al. \citeyear{Urpo1997}), precisely the range currently lacking in ALMA observations. For this study, we utilize Mets\"ahovi full-disk observations of a PR at a wavelength of 8.10 mm (beam size = 144") observed between May 27, 1993 and July 16, 1994. The PR brightness temperatures measurements from Mets\"ahovi were obtained using a method analogous to that described for ALMA, but applied to the full-disk image with the limb brightening effect already corrected. The resulting Mets\"ahovi measurements, including the corresponding QS brightness temperature measurements, are shown in Table \ref{Table_data}. Removing the limb brightening effect in advance gives $T_\mathrm{b}$(QS) = $T_\mathrm{b}$(QS$_\mathrm{central}$) in Table \ref{Table_data} for Mets\"ahovi data. Consequently, the derived $\Delta T_\mathrm{b}$(QS) value reflects the general comparison of the brightness temperature between the PR and QS as if they were already in the center of the solar disk, far away from the effects of limb brightening.

In addition to Mets\"ahovi data, we also incorporated the PR brightness temperature measurements previously obtained with the 45-m diameter parabolic radio telescope at the Nobeyama Radio Observatory\footnote{\url{https://www.nro.nao.ac.jp/en/}}. The measurements were adopted from Hiei et al. (\citeyear{Hiei1986}), where a PR structure was observed between July 16 and 22, 1984 at wavelengths of 3.10 mm (beam size = 17") and 8.30 mm (beam size = 46"). The brightness temperature of the PR structure in Hiei et al. (\citeyear{Hiei1986}) at these two wavelengths was determined by measuring the depression in the radio emission observed by the Nobeyama 45-m radio telescope using the already known brightness temperature value of the QS region at the same wavelengths. The said PR brightness temperature measurements and the corresponding QS measurements are listed in Table \ref{Table_data}. As with the Mets\"ahovi data, we also note that the effect of limb brightening has already been accounted for in the Nobeyama image data. So, for the measurements from all three radio instruments (ALMA, Mets\"ahovi, and Nobeyama) the center-to-limb brightening correction was taken into account.

\section{Modeling results and comparison with observations}
\label{Results}
Following the modeling procedure described in Section \ref{Calculation}, we obtained the best-fitted PR brightness temperature profiles by using the modified ATLCW QS model as the input atmospheric model for our calculations. The results are presented in the next two subsections, where we analyze the PR on the solar disk under two different assumptions regarding its stability: hydrostatic equilibrium and non-hydrostatic equilibrium.

We should note that, similar to Paper \Romannum{1}, we conducted our modeling analysis on two different sets of measurements. The first set comprises actual measurements corresponding to the sum of the brightness temperature of the measured central QS region and $\Delta T_\mathrm{b}$ from Table \ref{Table_data}. These measurements and the later models are represented with red symbols and the index "a". The second set, referred to as differential measurements, is obtained using a similar procedure by now adding $\Delta T_\mathrm{b}$ from Table \ref{Table_data} to the QS brightness temperature predicted by the brightness temperature profile obtained for the original and unperturbed ATLCW QS model based on the procedure described in Section \ref{Calculation} for each wavelength. These differential measurements and their associated models are represented with blue symbols and the index "b". This approach, as in Paper \Romannum{1}, is used to assess the sensitivity of the modeling results to small variations in brightness temperature measurements.

\subsection{Prominence (PR) in hydrostatic equilibrium}
\label{prominence_result1}
\begin{figure}[h!]
\centering
\resizebox{0.99\hsize}{!}{\includegraphics{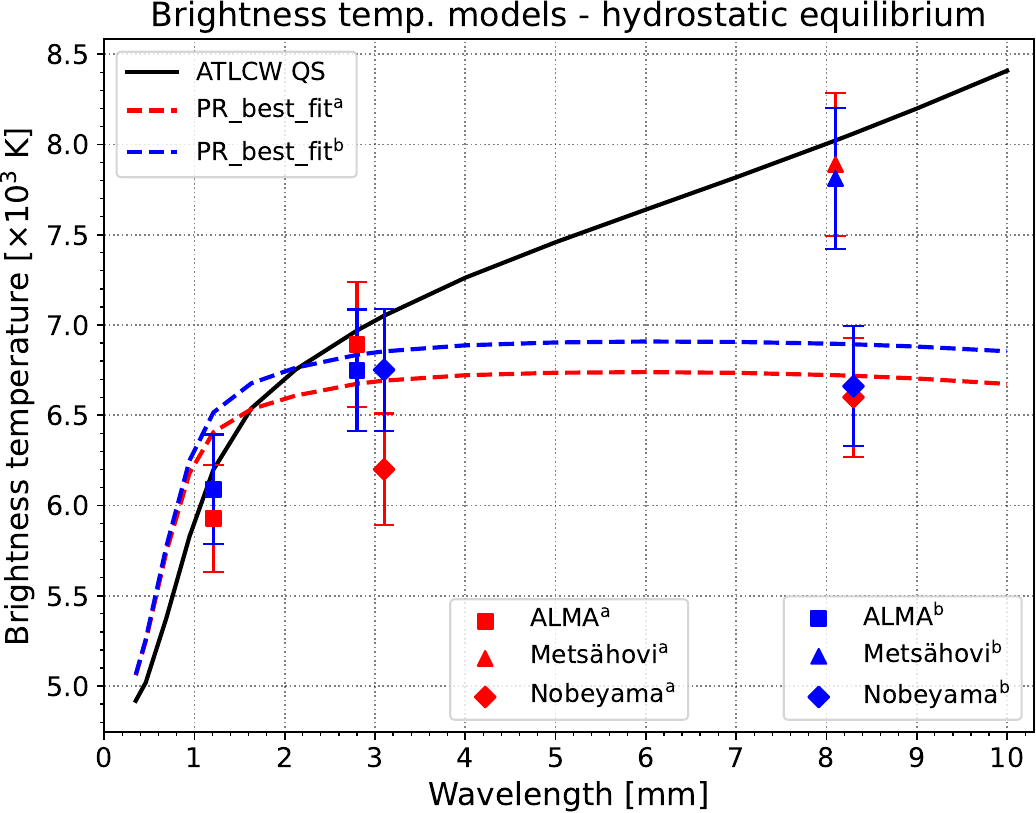}}
\caption{~Calculated brightness temperature profile obtained with the original, unperturbed ATLCW QS model (solid black curve), together with the best-fit PR brightness temperature profiles (PR$\_$best$\_$fit) obtained by fitting the perturbed ATLCW QS model to the ALMA, Mets\"ahovi, and Nobeyama measurements obtained using measurement procedures "a" (red symbols) and "b" (blue symbols). The error bars for all measurements correspond to the assumed 5\% measurement error for radio observations.}
\label{PR_hyd_eq}
\end{figure}

Based on the brightness temperature measurements discussed in Section \ref{observational_results}, Figure \ref{PR_hyd_eq}, as well as Figure \ref{PR_non_hyd_eq}, show the real (red) and differential (blue) ALMA (square symbol) and Mets\"ahovi measurements (triangular symbol) for PR slightly below the reference QS brightness temperature profile (solid black curve) obtained using the original, unperturbed ATCLW QS model. This indicates a small difference in emission intensity between PR and QS, with the PR structure appearing slightly darker compared to the surrounding QS. This is consistent with the actual observations of the emission of the PR structure in different parts of the electromagnetic spectrum, where the PR structure appears as a dark feature on the solar disk.

Compared to Mets\"ahovi, the Nobeyama measurements (red and blue diamond symbols for the actual and differential measurements, respectively) in Figures \ref{PR_hyd_eq} and \ref{PR_non_hyd_eq} show even lower PR brightness temperature values than Mets\"ahovi at a similar wavelength. The greater contrast between PR and QS observed between Mets\"ahovi and Nobeyama may be attributed to the superior spatial resolution of Nobeyama single-dish observations. Unlike Mets\"ahovi, the Nobeyama measurements are therefore less affected by the beam-convolution effect (e.g., Vr{\v s}nak et al. \citeyear{Vrsnak1992}), which can otherwise reduce contrast by blending radiation from adjacent regions into the measurement of the target structure. Lower spatial resolution effectively dilutes the measured emission of the PR by including contributions from its surroundings, thereby diminishing the observed brightness temperature difference. This effect is particularly pronounced at wavelengths around 8 mm, where Mets\"ahovi measurement shows a substantially smaller contrast relative to the QS level than Nobeyama, a discrepancy that correlates with the notable difference in the beam area size by a factor of 9.8 between the two instruments at this wavelength (see Table \ref{Table_data}).

It should be noted that ALMA, Mets\"ahovi, and Nobeyama all observed a different PR structure. Consequently, it is possible that a particular PR observed by Nobeyama exhibits significantly different physical properties than the PR observed by Mets\"ahovi at a similar wavelength, which could account for the differences in observed brightness temperature.

In the case of hydrostatic equilibrium (Figure \ref{PR_hyd_eq}), we obtained brightness temperature profiles for the PR structures (dashed curves) that closely follow the ALMA and Nobeyama measurements over the entire ALMA wavelength range. This agreement holds for both measurement procedures, "a" and "b", with procedure "b" (dashed blue curve in Figure \ref{PR_hyd_eq}) yielding slightly higher brightness temperatures by almost the same value (less than 250 K) compared to the profile based on measurement procedure "a" (red dotted curve in Figure \ref{PR_hyd_eq}). This offset begins at wavelengths shorter than 2 mm and persists throughout the ALMA range. What both measurement procedures have in common is that they result in brightness temperature profiles that show a progressively increasing contrast in brightness temperature between the PR structure and the QS level, which is in line with expectations. While most measurements do not align perfectly with the modeled PR profiles, the majority fall within their respective uncertainties relative to the predicted PR profile. The only notable outlier is the Mets\"ahovi measurement at a wavelength of 8.1 mm, which deviates significantly from the modeled PR profile.

Although all measurements lie below the QS profile, the modeled PR profiles exhibit a slight but significant increase in brightness temperature above the QS values predicted by the original ATLCW QS model below 2 mm wavelength. This is inconsistent with expectations based on observational data, where the PR should appear darker than the surrounding QS region. Furthermore, the modeled PR profiles display a maximum in the wavelength range of 4 to 6 mm, followed by a gradual and continuous decline in brightness temperature at longer wavelengths. This behavior is unexpected and lacks physical plausability, as solar structures generally exhibit some increase in temperature with height in the solar atmosphere. For cool structures such as the PR, temperatures at larger heights should at least remain comparable to those at lower atmospheric layers. While the overall PR profiles generally align with the observed measurements, these unphysical features, namely the excess brightness at short wavelengths and the unrealistic temperature decline at longer wavelengths, cast doubt on the validity of the hydrostatic equilibrium assumption as the primary stability mechanism for the PR structure.

The resulting multiplicative density factors ($f^\mathrm{a,b}_n$) and temperature factors ($f^\mathrm{a,b}_T$) corresponding to the best-fitted modeled brightness temperature profiles of the PR structure under the assumption of hydrostatic equilibrium are presented in Table \ref{Table_2} under the label PR1\_fit. For the density factors ($f^\mathrm{a,b}_n$ best fit for PR1\_fit in Table \ref{Table_2}), both measurement procedures yield consistent results, indicating that the density of the PR structure in the chromospheric region is approximately 159 -- 163 times greater than that of the surrounding QS. In the context of hydrostatic equilibrium, the temperature factor ($f^\mathrm{a,b}_T$ best fit for PR1\_fit in Table \ref{Table_2}) is simply the inverse of the density factor. This implies that the plasma temperature within the PR structure is reduced by the same factor relative to the QS temperature in the chromosphere. These values of the multiplicative density and temperature factors are consistent with the expected values of the physical parameters within the hydrostatic model for the PR structure (Parenti \citeyear{Parenti2014}).
\begin{table*}[h!]
\centering
\caption{~Output multiplicative factors for electron density ($f_n$ best fit) and temperature ($f_T$ best fit) corresponding to the best-fitting model for two measurement procedures ("a" and "b") calculated for PR structures with the corresponding minimum $\chi^2_\mathrm{min}$ value determined from the $\chi^2$-minimization method. The $f_n$ range and $f_T$ range correspond to the input range of values of the multiplicative density and temperature factor (Equation \ref{multi_factors}), within which the values of the two factors are varied to globally modify the density and temperature parameters of the input atmospheric model within a given perturbation height range until a global $\chi^2_\mathrm{min}$ value is found. The integration height range is also given, within which the integrating calculation of the brightness temperature was performed. A detailed description of the fitting procedure and the factor calculation can be found in Section \ref{Calculation}.}
\label{Table_2} 
\centering
\resizebox{2\columnwidth}{!}{
\begin{tabular}{c c c c c c c | c c c c}
\hline
\multicolumn{7}{c|}{Input parameters}&\multicolumn{4}{c}{Output parameters}\\
\midrule
PR stability&Atm. model &Integr. height range& Perturb. height range & $f_n$ range & $f_T$ range &Procedure& $f_n$ best fit & $f_T$ best fit  & $\chi^2_{\mathrm{min.}}$ & Reference\\
(condition)&(symbol) &(km)& (km) & (number) & (number) & (index)& (number) &(number)  &(number) & (symbol)\\
\hline\midrule
Hydrostatic&\multirow{2.5}{*}{ATLCW QS}    &\multirow{2.5}{*}{$0-57\;797$}&  \multirow{2.5}{*}{$40\;000-50\;000$}  & \multirow{2.5}{*}{$1-300$}&\multirow{2.5}{*}{$=1/\left(f_n\text{ range}\right)$}&a& $163^{+4}_{-4}$ & $=1/\left(f_n \text{ best fit}\right)$& $14.39$  & \multirow{2.5}{*}{PR1\_fit}\\[0.5em]
equilibrium&&&&&  &b&  $159^{+4}_{-4}$ &  $=1/\left(f_n \text{ best fit}\right)$& $8.11$ & \\ \cmidrule{1-11}
Non-hydrostatic&\multirow{2.5}{*}{ATLCW QS}&\multirow{2.5}{*}{$0-57\;797$}&  \multirow{2.5}{*}{$40\;000-50\;000$} &\multirow{2.5}{*}{$1-300$}& \multirow{2.5}{*}{$1/300-1$} & a&  $68^{+35}_{-23}$ & $1/\left(159^{+5}_{-4}\right)$ &  $11.69$ &\multirow{2.5}{*}{PR2\_fit} \\[0.5em]
equilibrium&&& && & b&  $60^{+43}_{-45}$ &  $1/\left(155^{+5}_{-4}\right)$ &  $5.84$ &\\ 
\midrule
\end{tabular}
}
\end{table*}

\subsection{Prominence (PR) in non-hydrostatic equilibrium}
\label{prominence_result2}
\begin{figure}[h!]
\centering
\resizebox{0.99\hsize}{!}{\includegraphics{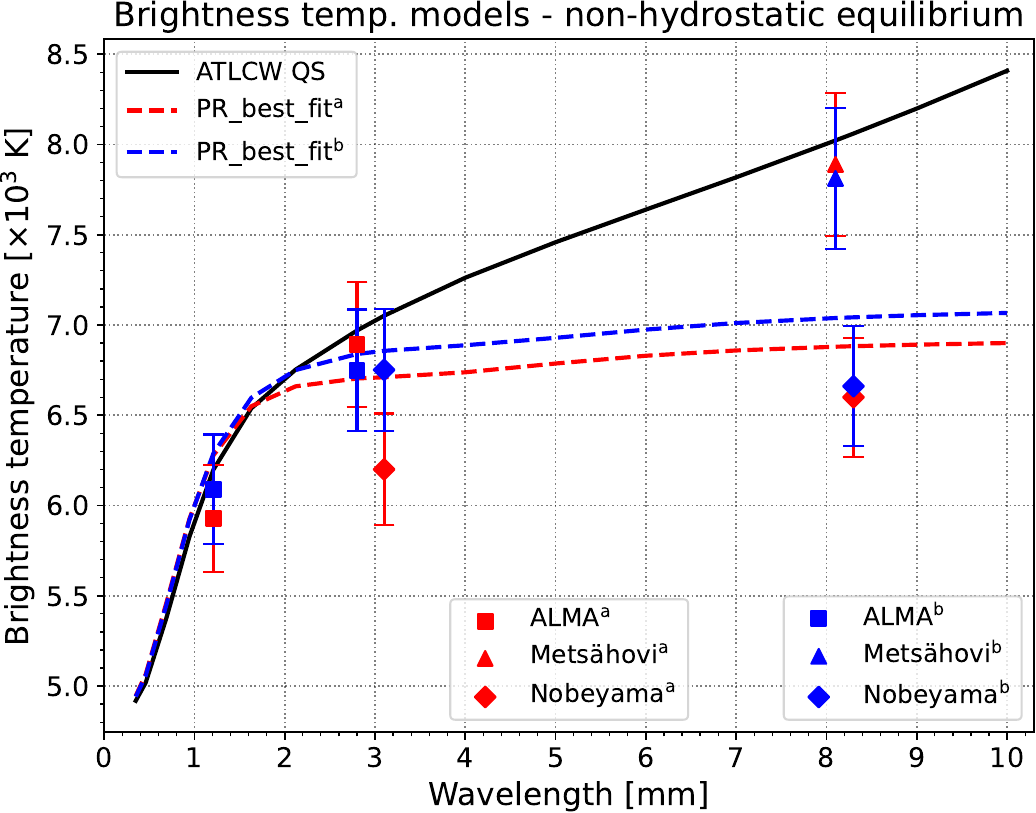}}
\caption{~Similar to Figure~\ref{PR_hyd_eq} using non-hydrostatic equilibrium assumption for the PR stability, where $f_n\neq1/f_T$.}
\label{PR_non_hyd_eq}
\end{figure}

In the case of non-hydrostatic equilibrium, where the hydrostatic equilibrium is not considered and more freedom is given to the input density and temperature factors, the resulting brightness temperature profiles (Figure \ref{PR_non_hyd_eq}) still show close agreement with the ALMA, Mets\"ahovi, and Nobeyama measurements, comparable to the results of the hydrostatic equilibrium scenario (Section \ref{prominence_result1}). Measurement procedures "a" and "b" again produced similar profiles, with procedure "b" yielding a profile (blue dashed curve in Figure \ref{PR_non_hyd_eq}) with slightly higher values, by less than 200 K higher, than the profile derived from procedure "a" (red dotted curve in Figure \ref{PR_non_hyd_eq}).

Both measurement procedures resulted in PR profiles with a significant continuous increase in the emission depression of the PR structure relative to the level of the surrounding QS region (black solid curve in Figure \ref{PR_non_hyd_eq}). This behavior is consistent with the expected dark appearance of the PR structure at high altitudes in the solar atmosphere. However, in contrast to the hydrostatic equilibrium case, the non-hydrostatic equilibrium scenario has a much more physically plausible behavior for the PR brightness temperature profiles at wavelengths shorter than 2 mm. In this case, the PR profiles not only align more closely with the measurements at these wavelengths, but they also nearly match the QS brightness temperature profile derived from the original ATLCW QS model. This suggests that there is no significant radiation emission contrast between the PR and the surrounding QS in the chromosphere at lower altitudes. This result is consistent with expectations, as both the PR structure and the surrounding QS region at these lower chromospheric altitudes share similar temperature and plasma density, leading to a comparable emission profile.

In addition, at wavelengths beyond 2 mm, the PR profiles show a slower, gradual increase in brightness temperature, with no significant change observed until the end of the ALMA wavelength range, compared to the value at about 2 mm, where the brightness temperature increase starts to slow down. This behavior is consistent with expectations, as the PR consists of trapped chromospheric plasma, which experiences minimal changes in temperature or density, and consequently, in brightness temperature, at higher altitudes in the solar atmosphere.

Compared to the previous hydrostatic case, the non-hydrostatic case here presents a more physical representation of the PR structure, aligning better with both the observations and the expected characteristics of the PR. This suggests that the stability of the PR structure is likely not governed by the hydrostatic equilibrium, but rather by another mechanism, most likely the PR magnetic field, which helps maintain the stability of the entire PR structure, even at coronal heights.

The resulting density ($f^\mathrm{a,b}_n$) and temperature factors ($f^\mathrm{a,b}_T$) corresponding to the best-fitted modeled PR brightness temperature profiles under the assumption that hydrostatic equilibrium does not apply are presented in Table \ref{Table_2} under the label PR2\_fit. In this non-hydrostatic case, the resulting temperature factor ($f^\mathrm{a,b}_T$ best fit for PR2\_fit in Table \ref{Table_2}) shows that the plasma temperature within the PR structure is 155 -- 159 times lower than the QS values. Given the uncertainties of the temperature factor, the values for the non-hydrostatic model are similar to those of the hydrostatic case, indicating that both cases yield comparable PR temperatures. However, a notable difference arises in the density factor ($f^\mathrm{a,b}_n$ best fit for PR2\_fit in Table \ref{Table_2}). In the non-hydrostatic case, the density factor is lower than in the hydrostatic case, indicating that the plasma density within the PR is 60 -- 68 times higher than in the surrounding QS. Accounting for uncertainties in the density factor, the non-hydrostatic case yields a maximum range for PR density that is 15 -- 103 times greater than the QS density.

As previously noted, based on the brightness temperature profiles in Figure \ref{PR_non_hyd_eq}, we visually estimated that the non-hydrostatic case presents a more physically plausible model compared to the hydrostatic case. This is further supported by the $\chi^2_\mathrm{min.}$ values (Table \ref{Table_2}), where the $\chi^2_\mathrm{min.}$ values for the non-hydrostatic case are significantly lower than those for the hydrostatic case, indicating a better agreement between the profiles and the actual observations. Moreover, the difference in $\chi^2_\mathrm{min.}$ between the hydrostatic and non-hydrostatic fit is highly statistically significant, indicating that it is rather unlikely that the improvement of the $\chi^2_\mathrm{min.}$ is just happening by chance. Therefore, the results strongly favor the non-hydrostatic model, which suggests that the plasma temperature within the PR is at least 150 times lower, and its plasma density is up to 100 times higher than those of the surrounding QS region in the chromosphere within the ALMA wavelength range. We should also note that the non-hydrostatic modeling had one more free parameter than the hydrostatic modeling in the fitting procedure, where both density and temperature of the ATLCW model were free and independent parameters. This gives an advantage to the non-hydrostatic modeling when it comes to producing a better fit, unlike the hydrostatic modeling which had only one free parameter out of the density and temperature, while the other was dependent on the free parameter via hydrostatic equilibrium condition.

\section{Discussion and conclusions}
\label{discussion_conclusion}
For the PR structures analyzed in this work, ALMA and Mets\"ahovi brightness temperature measurements showed that the PR on the disk has a brightness temperature slightly lower than the measured brightness temperature of the surrounding QS region (Table \ref{Table_data}), but also slightly lower than the brightness temperature profile predicted by the QS profile of the original ATLCW QS model (black solid curve in Figures \ref{PR_hyd_eq} and \ref{PR_non_hyd_eq}). On the other hand, the Nobeyama measurements revealed a significantly lower brightness temperature of the PR structure when compared to both the measured QS region and the QS profile derived from the original ATLCW QS model.

As discussed in Section \ref{prominence_result1}, the greater contrast between the Nobeyama measurements and the QS level is likely due to the beam-convolution effect, which is more pronounced in Mets\"ahovi measurements. Since Mets\"ahovi observations have a lower spatial resolution compared to the Nobeyama 45-m radio telescope at a similar wavelength, the reduced resolution generally leads to a lower contrast in the radiation emission between the observed structure and the surrounding regions due to inclusion of some part of the radiation from regions adjacent to the observed structure in the measured average emission. This effect can be seen in the positioning of the Mets\"ahovi and Nobeyama PR measurements relative to the QS brightness temperature profile derived from the original ATLCW QS model. If we consider the average H$\alpha$ width of about 25" of the PR studied in Braj{\v s}a et al. (Brajsa2018b) and in the present work (Figure \ref{PR_Halpha_ALMA_image}), about 5.8 PR widths would fit within the Mets\"ahovi beam size, while only about 1.8 PR widths would fit within the Nobeyama beam size, clearly showing that the obtained Mets\"ahovi PR measurements are under the influence of the beam-convolution effect.

Using the same procedure of the present work, a separate modeling analysis was performed with exclusion of the Mets\"ahovi measurements. Compared to the current work, the results without the Mets\"ahovi measurements included showed a decrease between 100 and 400 K in the brightness temperature of the long-wavelength tail of the best-fitted brightness temperature profile for both hydrostatic and non-hydrostatic cases. The short-wavelength part of the brightness temperature profiles, especially below 2 mm, showed no significant difference from our models that include Mets\"ahovi measurements. The vertical fall of the profile tail at longer mm wavelengths was due to a slight increase in the density and decrease in temperature factors for hydrostatic case and a slight decrease in density and temperature factors for non-hydrostatic case, all within uncertainties of the presented results of this work. 

To assess how spatial resolution and beam-convolution effects influence the measured brightness temperature and, consequently, the models, the use of interferometric ALMA observations should be considered. Using such data to model the PR brightness temperature would help determine whether the increased spatial resolution of interferometric observations, compared to the single-dish data used in this work, leads to significant changes in both the measurements and the resulting models. However, the interferometric ALMA data have certain limitations. The first is the small field of view, which may not encompass the entire PR structure, necessitating the use of a mosaic observing method. Second, the currently available ALMA data have a limited time window for observations that are not specifically targeted at PR structures, meaning that sporadic PRs would need to be identified within the available data. This limitation reduces observational opportunities and introduces potential biases. We plan to incorporate interferometric ALMA data for PR modeling in the future, as more solar ALMA data become available, thereby increasing our chances of capturing PRs observed in interferometric mode. Regardless of this, the brightness temperature measurements from all instruments indicate that the PR on the solar disk has a lower brightness temperature compared to the surrounding QS region over the entire ALMA wavelength range.

Certain ALMA limitations will be handled in future projects, especially the small number of currently observing wavelength bands available for solar analysis. The current main goal of future ALMA solar observations is to include more wavelength bands and upgrade on the existing bands to get a better coverage of the ALMA range. The currently used ALMA Band 3 ($\lambda\approx3$ mm) and Band 6 ($\lambda\approx1.2$ mm) observations cover the solar atmosphere in the height range of 600 -- 1\;600 km and 400 -- 1\;400 km, respectively, with the highest sensitivity at around 960 km for Band 3 and at around 730 km for Band 6 (Figure 5 in Wedemeyer et al. \citeyear{Wedemeyer2016}). Future ALMA bands will cover even longer wavelengths up to around 9 mm, extending the coverage to well above 1\;700 km, with the highest sensitivity at around 1\;100 to 1\;200 km (Figure 5 in Wedemeyer et al. \citeyear{Wedemeyer2016}). Another important step in the future will be the inclusion of the newly planned 50-m single-dish Atacama Large Aperture Submillimeter Telescope (AtLAST) radio telescope (e.g, Booth et al. \citeyear{Booth2024}; Wedemeyer et al. \citeyear{Wedemeyer2024}). AtLAST would compensate with simultaneous wider frequency coverage, thus unlocking complementary science cases and boosting the impact of mm and sub-mm observations on solar physics beyond its current state.

Moreover, the PR brightness temperature profile results of our work for both the hydrostatic and non-hydrostatic cases exhibit a similar behavior, with profiles from both cases agreeing reasonably well with most PR brightness temperature measurements, regardless of the assumption of (non-)hydrostatic equilibrium. Additionally, the hydrostatic and non-hydrostatic cases yielded density and temperature factors (PR1\_fit and PR2\_fit in Table \ref{Table_2}) indicating a significantly higher density and lower plasma temperature within the PR structure compared to the surrounding QS region. This result is consistent with the expected physical properties of PRs relative to QS and demonstrates that the PR structure contains denser and cooler plasma than the surrounding solar atmosphere, not only in the corona but also at chromospheric heights.

Although the brightness temperature profiles for the PR structure in both the hydrostatic and non-hydrostatic cases exhibit a similar appearance and behavior, there are two key differences between the models for these cases. The first and most notable difference is observed in the behavior of the PR brightness temperature profiles relative to the QS profile derived from the original ATLCW QS model. For wavelengths longer than 2 mm, both the hydrostatic and non-hydrostatic cases show that the PR brightness temperature is significantly lower than that of the QS, with the difference increasing at longer wavelengths. This trend is consistent with the expected increase in the emission contrast between the PR and the surrounding QS as the height in the solar atmosphere increases.

At wavelengths shorter than 2 mm, however, the hydrostatic case shows considerably higher brightness temperature than the QS region, whereas the non-hydrostatic case remains nearly at the QS level. In this range, the hydrostatic case deviates from the expected physical behavior of the PR, where we expect a PR would be darker than the surrounding QS. The greater physical realism of the non-hydrostatic case is further supported by its better agreement with all brightness temperature measurements, indicating a better fit for the non-hydrostatic equilibrium case. These results clearly favor the non-hydrostatic model, implying that the stability within the PR structure is most likely not governed by hydrostatic equilibrium, but rather by some other mechanism, most likely related to the magnetic field of the PR structure. We should note that the good fit between the calculated and measured brightness temperatures also confirms that the previously assumed thermal bremsstrahlung is indeed the dominant radiation mechanism of PRs in the mm and sub-mm wavelength range, as was the case for QS, AR, and CH structures studied in Paper \Romannum{1}.

The second difference between the non-hydrostatic and hydrostatic cases is evident in the density and temperature factors ($f_n$ best fit and $f_T$ best fit for PR1\_fit and PR2\_fit in Table \ref{Table_2}). While both cases show similar temperature factors, with the plasma temperature within the PR structure being more than 150 times lower than the QS temperature, the hydrostatic and non-hydrostatic cases lead to significantly different density factor values. In particular, the hydrostatic case results in a PR density that is about 160 times higher than the QS density, whereas the non-hydrostatic case results in a density that is only 60 -- 70 times higher than the QS density. Considering the optical depth for the thermal bremsstrahlung emission given in Equation (\ref{Eq_optical_depth}) and similarity in the temperature factors, the non-hydrostatic case predicts less dense and optically thinner PRs than the hydrostatic case. The same is obtained even without Mets\"ahovi measurements as possible outliers.

However, the profiles for both the hydrostatic and non-hydrostatic cases exhibit very similar appearances regardless of the differences in the density factors. Therefore, a significant change in the PR plasma density, as observed between the hydrostatic and non-hydrostatic cases, may not lead to a significant change in the PR brightness temperature profile. The weak density dependence of the PR brightness temperature is further highlighted by the large uncertainty in the density factors, indicating that, for a given PR temperature, a wide range of densities (15 --103 times that of the QS values) leads to a similar brightness temperature. This implies that the PR brightness temperature is primarily influenced by the PR plasma temperature, with the significantly lower PR temperature compared to the QS causing the visible contrast in radiation emission between the PR and the QS. This finding further supports the conclusion that the hydrostatic case is not a suitable model for the PR because hydrostatic equilibrium would require that significant changes in density lead to corresponding changes in temperature, which is not observed in this study.

If we assume that the PR pressure equilibrium is not hydrostatic, nor hydrodynamic, but is instead maintained through the magnetic field permeating the PR, as our results suggest, we can calculate the excess average magnetic field within the PR against the surrounding QS magnetic field by using the equilibrium relation between the sums of plasma and magnetic pressure of the QS and PR given, in cgs system, as (e.g., Benz \citeyear{Benz2002}):
\begin{equation}
\label{PR_eq1}
2n_\mathrm{QS}k_\mathrm{B}T_\mathrm{QS}+\frac{B^2_\mathrm{QS}}{8\pi}=2n_\mathrm{PR}k_\mathrm{B}T_\mathrm{PR}+\frac{B^2_\mathrm{PR}}{8\pi},
\end{equation}
where $B^2_\mathrm{QS}$ and $B^2_\mathrm{PR}$ are the average QS and PR magnetic fields, respectively, $k_\mathrm{B}$ is the Boltzmann constant, $n_\mathrm{QS}$ and $n_\mathrm{PR}$ are the QS and PR (electron) densities, and $T_\mathrm{QS}$ and $T_\mathrm{PR}$ are the QS and PR (electron) temperatures, respectively. The numerical factor 2 is the correction factor which includes the contributions from both electrons and ions, which here contribute separately to the overal plasma pressure with equal values. Through the parameter manipulation, the Equation (\ref{PR_eq1}) can be written as:
\begin{equation}
\label{PR_eq2}
16\pi k_\mathrm{B}n_\mathrm{QS}T_\mathrm{QS}\times(1-\frac{n_\mathrm{PR}}{n_\mathrm{QS}}\times\frac{T_\mathrm{PR}}{T_\mathrm{QS}})=B^2_\mathrm{PR}-B^2_\mathrm{QS}\approx(\Delta B)^2,
\end{equation}
where $\Delta B$ is the average additional magnetic field in the PR, that is, the excess field compared to the surrounding QS field, while fractions $\frac{n_\mathrm{PR}}{n_\mathrm{QS}}=f_n$ and $\frac{T_\mathrm{PR}}{T_\mathrm{QS}}=f_T$ correspond to the multiplicative density and temperature factors, respectively, which we obtained from our modeling in Table \ref{Table_2}. We should note that the final step in Equation \ref{PR_eq2} is valid only if the magnetic field of the quiet corona is much less than the magnetic field of the PR itself. According to recent measurements, the quiet-corona magnetic field at PR heights is indeed small with values of only a small fraction of 1 G (Yang et al. \citeyear{Yang2024}). Assuming the best fitted density and temperature factors PR2\_fit from Table \ref{Table_2} for the non-hydrostatic equilibrium model and considering the coronal PR regions with the surrounding QS corona having $n_\mathrm{QS}=2.3\times10^8$ cm$^{-3}$ and $T_\mathrm{QS}=1.1\times10^6$ K, we get $\Delta B=1$ G. This value of the PR magnetic field is of the similar order of magnitude as PR models calculated in earlier works like Engvold et al. (\citeyear{Engvold1990}), Jensen \& Wiik (\citeyear{Jensen1990}), and Bommier et al. (\citeyear{Bommier1994}), who give magnetic fields for quiescent PRs from 2 to 20 G, with a preference towards lower values.

Considering the PR density range of $10^9-10^{11}$ cm$^{-3}$ given by Parenti (\citeyear{Parenti2014}) and the average density from the ATLCW QS model (Table 1 in Avrett et al. \citeyear{Avrett2015}) within the height range of $40\;000-50\;000$ km used for the modeling, the resulting density factors ($f_n$ best fit for PR1\_fit and PR2\_fit in Table \ref{Table_2}), along with the uncertainties in the results, yield PR densities in the range of about $0.35-3.9\times 10^{10}$ cm$^{-3}$. These PR density values derived from our density factors agree well with the values reported by Parenti (\citeyear{Parenti2014}). Good agreement is also found with much older studies, such as, Engvold et al. (\citeyear{Engvold1990}) and Jensen \& Wiik (\citeyear{Jensen1990}). These two studies give a similarly wide range of values for the electron density in quiescent PRs of $10^{10}-10^{11}$ cm$^{-3}$ in the central part and about $4\times10^9$ cm$^{-3}$ at the edge of the PR structure. Our findings overlap well with the full range (including the center and the edge of a PR) of the PR density values reported by these earlier studies.

On the other hand, Parenti (2014) estimates a PR temperature range of $7\;500-9\;000$ K, while Okada et al. (\citeyear{Okada2020}) report temperatures between 8\;000 and 12\;000 K. Using the average temperature provided in the ATLCW QS model (Table 1 in Avrett et al. \citeyear{Avrett2015}) within the same considered PR height range, including the results uncertainties, the temperature factors ($f_T$ best fit for PR1\_fit and PR2\_fit in Table \ref{Table_2}) give PR temperatures in the range of $6\;290-6\;957$ K. These values are below the minimum values reported by Okada et al. (\citeyear{Okada2020}) and Parenti (\citeyear{Parenti2014}). However, our results are consistent with earlier PR temperature measurements listed in Table 1 of Okada et al. (\citeyear{Okada2020}), which span a range of $4\;000-20\;000$ K. Our results fall on the lower side of this temperature range, suggesting that the PRs observed in our study are of the cooler type. A similar agreement is also found with two previously mentioned studies Engvold et al. (\citeyear{Engvold1990}) and Jensen \& Wiik (\citeyear{Jensen1990}), which report temperatures in the range of $4\;300-8\;500$ K for the central part and $8\;000-12\;000$ K at the edge of a PR, again showing a wide range of PR temperatures. Our results fall well within the PR temperature range reported in these earlier studies.

Compared to previous PR modeling studies such as Braj{\v s}a et al. (\citeyear{Brajsa2009}, \citeyear{Brajsa2018a}), the modeling method used in the present work demonstrates improvements in two key areas. First, unlike the two earlier studies, which first constructed a model with assumed changes in density and temperature parameters and then compared it with QS models and/or real PR observations to see which model best matches expectations, our approach first fits the input atmospheric model using varying density and temperature parameters. This process leads to a model that more accurately aligns with our expectations. Second, our work employs a newer and more physically realistic atmospheric model for the QS as the basis for PR modeling, as demonstrated in Paper \Romannum{1}. This contrasts with the older FAL model used in the previous studies, which is less reliable at longer mm wavelengths.

Finally, based on the results from modeling the brightness temperature of PR structures, it is clear that using the ATLCW QS atmospheric model as the foundation for PR models is adequate for physically describing PR properties, with the obtained models agreeing very well with our and previous measurements. Similar to the brightness temperature models for QS, AR, and CH regions analyzed in Paper \Romannum{1}, which showed excellent agreement with observations, the models presented here for PRs are similarly well-matched to the observational data. Moreover, thermal bremsstrahlung was assumed to be the dominant radiation mechanism in our brightness temperature calculations, consistent with the approach used in Paper \Romannum{1}. Thus, based on the well-fitted PR brightness temperature profiles presented in this work, we can conclude that the mm and sub-mm emission measured by ALMA and other radio telescopes in the ALMA wavelength range for PRs primarily originates from thermal bremsstrahlung, just as it does for the QS, AR, and CH regions studied in Paper \Romannum{1}.

In a future work, we plan to extend the current 1D semi-empirical modeling of PRs to a 2D and, if possible, eventually to 3D models. This extension naturally follows from the height-variation 1D modeling of PRs discussed earlier and builds upon the framework presented in this work. By advancing to higher dimensions, we would be able to measure the brightness temperature at each location (e.g., each pixel) within the structure, where the modeling procedure outlined in this paper would then be applied to each of these locations. The output of such approach would be a 2D map of the brightness temperature, along with corresponding maps of the density and temperature of the PR structure at various observed wavelengths (i.e., heights in the atmosphere). This approach would enable the study of cross-sections of the Sun's atmospheric structure and provide insights into brightness temperature variations and other physical properties within different layers of the observed structure. Extending this 2D modeling to three dimensions would involve stacking the modeled 2D layers along the third dimension (i.e., height), creating a more comprehensive model of the PR structure. The described model extension will be the subject of our future research efforts.

\section*{Acknowledgments}

This work was supported by the Croatian Science Foundation as part of the "Young Researchers' Career Development Project - Training New Doctorial Students" under the project 7549 "Millimeter and submillimeter observations of the solar chromosphere with ALMA". Support from the Austrian-Croatian Bilateral Scientific Projects ”Comparison of ALMA observations with MHD-simulations of coronal waves interacting with coronal holes” and ”Multi-Wavelength Analysis of Solar Rotation Profile” is also acknowledged. It has also received funding from the Horizon 2020 project SOLARNET (824135, 2019–2023). In this paper, ALMA data ADS/JAO.ALMA\#2011.0.00020.SV were used. ALMA is a partnership of ESO (representing its member states), NSF (USA) and NINS (Japan), together with NRC (Canada), MOST and ASIAA (Taiwan), and KASI (Republic of Korea), in cooperation with the Republic of Chile. The Joint ALMA Observatory is operated by ESO, AUI/NRAO, and NAOJ. We thank the ALMA project for enabling solar observations with ALMA. This publication also uses data from the Mets\"ahovi Radio Observatory, operated by the Aalto University (Aalto University \citeyear{Aalto2019}). The work also utilizes GONG data acquired by instrument operated by the Cerro Tololo Interamerican Observatory, which was obtained by the NSO Integrated Synoptic Program (NISP), managed by the National Solar Observatory, the Association of Universities for Research in Astronomy (AURA), Inc. under a cooperative agreement with the National Science Foundation. RB acknowledges financial support from the Alexander von Humboldt Foundation. CLS acknowledges financial support from the S{\~a}o Paulo Research Foundation (FAPESP), grant number 2019/03301-8.

\appendix
\bibliography{Matkovic_et_al_prominence_brightness_temp_model}

\begin{thebibliography}{}

\bibitem [\protect \citeauthoryear {%
{Aalto~University}%
}{%
{Aalto~University}%
}{%
{\protect \APACyear {2019}}%
}]{%
Aalto2019}
\APACinsertmetastar {%
Aalto2019}%
\begin{APACrefauthors}%
{Aalto~University}.%
\end{APACrefauthors}%
\unskip\
\newblock
\APACrefYearMonthDay{2019}{}{},
\newblock
\APACrefbtitle {Mets{\"a}hovi Radio Observatory public solar database, Aalto
  University, Mets{\"a}hovi Radio Observatory.} {Mets{\"a}hovi Radio
  Observatory public solar database, Aalto University, Mets{\"a}hovi Radio
  Observatory.},
\newblock
\APAChowpublished
  {\url{http://urn.fi/urn:nbn:fi:att:f371cb6d-f84c-4d76-99e4-c39c639fd0de}}.
\PrintBackRefs{\CurrentBib}

\bibitem [\protect \citeauthoryear {%
{Alissandrakis}%
, {Bastian}%
\BCBL {}\ \BBA {} {Nindos}%
}{%
{Alissandrakis}%
\ \protect \BOthers {.}}{%
{\protect \APACyear {2022}}%
}]{%
Alissandrakis2022}
\APACinsertmetastar {%
Alissandrakis2022}%
\begin{APACrefauthors}%
{Alissandrakis}, C\BPBI E.%
, {Bastian}, T\BPBI S.%
\BCBL {}\ \BBA {} {Nindos}, A.%
\end{APACrefauthors}%
\unskip\
\newblock
\APACrefYearMonthDay{2022}{}{},
\newblock
\unskip
\newblock
\APACjournalVolNumPages{\aap}{661}{}{L4}.
\PrintBackRefs{\CurrentBib}

\bibitem [\protect \citeauthoryear {%
{Avrett}%
, {Tian}%
, {Landi}%
, {Curdt}%
\BCBL {}\ \BBA {} {W{\"u}lser}%
}{%
{Avrett}%
\ \protect \BOthers {.}}{%
{\protect \APACyear {2015}}%
}]{%
Avrett2015}
\APACinsertmetastar {%
Avrett2015}%
\begin{APACrefauthors}%
{Avrett}, E.%
, {Tian}, H.%
, {Landi}, E.%
, {Curdt}, W.%
\BCBL {}\ \BBA {} {W{\"u}lser}, J\BPBI P.%
\end{APACrefauthors}%
\unskip\
\newblock
\APACrefYearMonthDay{2015}{}{},
\newblock
\unskip
\newblock
\APACjournalVolNumPages{\apj}{811}{}{87}.
\PrintBackRefs{\CurrentBib}

\bibitem [\protect \citeauthoryear {%
{Bastian}%
, {Ewell}%
\BCBL {}\ \BBA {} {Zirin}%
}{%
{Bastian}%
\ \protect \BOthers {.}}{%
{\protect \APACyear {1993}}%
}]{%
Bastian1993}
\APACinsertmetastar {%
Bastian1993}%
\begin{APACrefauthors}%
{Bastian}, T\BPBI S.%
, {Ewell}, M\BPBI W., Jr.%
\BCBL {}\ \BBA {} {Zirin}, H.%
\end{APACrefauthors}%
\unskip\
\newblock
\APACrefYearMonthDay{1993}{}{},
\newblock
\unskip
\newblock
\APACjournalVolNumPages{\apj}{418}{}{510}.
\PrintBackRefs{\CurrentBib}

\bibitem [\protect \citeauthoryear {%
{Benz}%
}{%
{Benz}%
}{%
{\protect \APACyear {2002}}%
}]{%
Benz2002}
\APACinsertmetastar {%
Benz2002}%
\begin{APACrefauthors}%
{Benz}, A\BPBI O.%
\end{APACrefauthors}%
\unskip\
\newblock
\APACrefYear{2002},
\newblock
\APACrefbtitle {{Astrophysics and Space Science Library. Vol. 279, Plasma
  Astrophysics (2nd ed., Dordrecht: Kluwer)}} {{Astrophysics and Space Science
  Library. Vol. 279, Plasma Astrophysics (2nd ed., Dordrecht: Kluwer)}}.
\PrintBackRefs{\CurrentBib}

\bibitem [\protect \citeauthoryear {%
{Benz}%
}{%
{Benz}%
}{%
{\protect \APACyear {2009}}%
}]{%
Benz2009}
\APACinsertmetastar {%
Benz2009}%
\begin{APACrefauthors}%
{Benz}, A\BPBI O.%
\end{APACrefauthors}%
\unskip\
\newblock
\APACrefYear{2009},
\newblock
\APACrefbtitle {{Radio Emission of the Quiet Sun. In: Tr\"umper J. (ed.), {\it
  Landolt B\"ornstein}, Springer-Verlag: Berlin, 103 -- 117}} {{Radio Emission
  of the Quiet Sun. In: Tr\"umper J. (ed.), {\it Landolt B\"ornstein},
  Springer-Verlag: Berlin, 103 -- 117}}.
\PrintBackRefs{\CurrentBib}

\bibitem [\protect \citeauthoryear {%
{Benz}%
, {Krucker}%
, {Acton}%
\BCBL {}\ \BBA {} {Bastian}%
}{%
{Benz}%
\ \protect \BOthers {.}}{%
{\protect \APACyear {1997}}%
}]{%
Benz1997}
\APACinsertmetastar {%
Benz1997}%
\begin{APACrefauthors}%
{Benz}, A\BPBI O.%
, {Krucker}, S.%
, {Acton}, L\BPBI W.%
\BCBL {}\ \BBA {} {Bastian}, T\BPBI S.%
\end{APACrefauthors}%
\unskip\
\newblock
\APACrefYearMonthDay{1997}{}{},
\newblock
\unskip
\newblock
\APACjournalVolNumPages{\aap}{320}{}{993-1000}.
\PrintBackRefs{\CurrentBib}

\bibitem [\protect \citeauthoryear {%
{Bommier}%
, {Landi Degl'Innocenti}%
, {Leroy}%
\BCBL {}\ \BBA {} {Sahal-Brechot}%
}{%
{Bommier}%
\ \protect \BOthers {.}}{%
{\protect \APACyear {1994}}%
}]{%
Bommier1994}
\APACinsertmetastar {%
Bommier1994}%
\begin{APACrefauthors}%
{Bommier}, V.%
, {Landi Degl'Innocenti}, E.%
, {Leroy}, J\BHBI L.%
\BCBL {}\ \BBA {} {Sahal-Brechot}, S.%
\end{APACrefauthors}%
\unskip\
\newblock
\APACrefYearMonthDay{1994}{}{},
\newblock
\unskip
\newblock
\APACjournalVolNumPages{\solphys}{154}{2}{231-260}.
\PrintBackRefs{\CurrentBib}

\bibitem [\protect \citeauthoryear {%
{Booth}%
\ \protect \BOthers {.}}{%
{Booth}%
\ \protect \BOthers {.}}{%
{\protect \APACyear {2024}}%
}]{%
Booth2024}
\APACinsertmetastar {%
Booth2024}%
\begin{APACrefauthors}%
{Booth}, M.%
, {Klaassen}, P.%
, {Cicone}, C.%
\ et al.\end{APACrefauthors}%
\unskip\
\newblock
\APACrefYearMonthDay{2024}{}{},
\newblock
\unskip
\newblock
\APACjournalVolNumPages{arXiv e-prints}{}{}{arXiv:2407.01413}.
\PrintBackRefs{\CurrentBib}

\bibitem [\protect \citeauthoryear {%
{Braj\v sa}%
}{%
{Braj\v sa}%
}{%
{\protect \APACyear {1993}}%
}]{%
Brajsa1993}
\APACinsertmetastar {%
Brajsa1993}%
\begin{APACrefauthors}%
{Braj\v sa}, R.%
\end{APACrefauthors}%
\unskip\
\newblock
\APACrefYearMonthDay{1993}{}{},
\newblock
\unskip
\newblock
\APACjournalVolNumPages{\solphys}{144}{}{199-202}.
\PrintBackRefs{\CurrentBib}

\bibitem [\protect \citeauthoryear {%
{Braj{\v s}a}%
\ \protect \BOthers {.}}{%
{Braj{\v s}a}%
\ \protect \BOthers {.}}{%
{\protect \APACyear {2007}}%
}]{%
Brajsa2007}
\APACinsertmetastar {%
Brajsa2007}%
\begin{APACrefauthors}%
{Braj{\v s}a}, R.%
, {Benz}, A\BPBI O.%
, {Temmer}, M.%
, {Jurdana-{\v S}epi{\'c}}, R.%
, {{\v S}aina}, B.%
\BCBL {}\ \BBA {} {W{\"o}hl}, H.%
\end{APACrefauthors}%
\unskip\
\newblock
\APACrefYearMonthDay{2007}{}{},
\newblock
\unskip
\newblock
\APACjournalVolNumPages{\solphys}{245}{}{167-176}.
\PrintBackRefs{\CurrentBib}

\bibitem [\protect \citeauthoryear {%
{Braj{\v s}a}%
\ \protect \BOthers {.}}{%
{Braj{\v s}a}%
\ \protect \BOthers {.}}{%
{\protect \APACyear {2018a}}%
}]{%
Brajsa2018a}
\APACinsertmetastar {%
Brajsa2018a}%
\begin{APACrefauthors}%
{Braj{\v s}a}, R.%
, {Kuhar}, M.%
, {Benz}, A\BPBI O.%
\ et al.\end{APACrefauthors}%
\unskip\
\newblock
\APACrefYearMonthDay{2018a}{}{},
\newblock
\unskip
\newblock
\APACjournalVolNumPages{Central European Astrophysical Bulletin}{42}{}{1}.
\PrintBackRefs{\CurrentBib}

\bibitem [\protect \citeauthoryear {%
{Braj{\v s}a}%
\ \protect \BOthers {.}}{%
{Braj{\v s}a}%
\ \protect \BOthers {.}}{%
{\protect \APACyear {2009}}%
}]{%
Brajsa2009}
\APACinsertmetastar {%
Brajsa2009}%
\begin{APACrefauthors}%
{Braj{\v s}a}, R.%
, {Rom{\v s}tajn}, I.%
, {W{\"o}hl}, H.%
, {Benz}, A\BPBI O.%
, {Temmer}, M.%
\BCBL {}\ \BBA {} {Ro{\v s}a}, D.%
\end{APACrefauthors}%
\unskip\
\newblock
\APACrefYearMonthDay{2009}{}{},
\newblock
\unskip
\newblock
\APACjournalVolNumPages{\aap}{493}{}{613-621}.
\PrintBackRefs{\CurrentBib}

\bibitem [\protect \citeauthoryear {%
{Braj{\v s}a}%
\ \protect \BOthers {.}}{%
{Braj{\v s}a}%
\ \protect \BOthers {.}}{%
{\protect \APACyear {2018b}}%
}]{%
Brajsa2018b}
\APACinsertmetastar {%
Brajsa2018b}%
\begin{APACrefauthors}%
{Braj{\v s}a}, R.%
, {Sudar}, D.%
, {Benz}, A\BPBI O.%
\ et al.\end{APACrefauthors}%
\unskip\
\newblock
\APACrefYearMonthDay{2018b}{}{},
\newblock
\unskip
\newblock
\APACjournalVolNumPages{\aap}{613}{}{A17}.
\PrintBackRefs{\CurrentBib}

\bibitem [\protect \citeauthoryear {%
{Engvold}%
, {Hirayama}%
, {Leroy}%
, {Priest}%
\BCBL {}\ \BBA {} {Tandberg-Hanssen}%
}{%
{Engvold}%
\ \protect \BOthers {.}}{%
{\protect \APACyear {1990}}%
}]{%
Engvold1990}
\APACinsertmetastar {%
Engvold1990}%
\begin{APACrefauthors}%
{Engvold}, O.%
, {Hirayama}, T.%
, {Leroy}, J\BPBI L.%
, {Priest}, E\BPBI R.%
\BCBL {}\ \BBA {} {Tandberg-Hanssen}, E.%
\end{APACrefauthors}%
\unskip\
\newblock
\APACrefYearMonthDay{1990}{}{},
\newblock
{\BBOQ}\APACrefatitle {{Hvar Reference Atmosphere of Quiescent Prominences}}
  {{Hvar Reference Atmosphere of Quiescent Prominences}}.{\BBCQ}
\newblock
\BIn{} V.~{Ruzdjak}\ \BBA {} E.~{Tandberg-Hanssen}\ (\BEDS), \APACrefbtitle
  {IAU Colloq. 117: Dynamics of Quiescent Prominences} {IAU Colloq. 117:
  Dynamics of Quiescent Prominences}\ \BVOL~363, \BPG~294.
\PrintBackRefs{\CurrentBib}

\bibitem [\protect \citeauthoryear {%
{Fontenla}%
, {Avrett}%
\BCBL {}\ \BBA {} {Loeser}%
}{%
{Fontenla}%
\ \protect \BOthers {.}}{%
{\protect \APACyear {1993}}%
}]{%
Fontenla1993}
\APACinsertmetastar {%
Fontenla1993}%
\begin{APACrefauthors}%
{Fontenla}, J\BPBI M.%
, {Avrett}, E\BPBI H.%
\BCBL {}\ \BBA {} {Loeser}, R.%
\end{APACrefauthors}%
\unskip\
\newblock
\APACrefYearMonthDay{1993}{}{},
\newblock
\unskip
\newblock
\APACjournalVolNumPages{\apj}{406}{}{319-345}.
\PrintBackRefs{\CurrentBib}

\bibitem [\protect \citeauthoryear {%
{Gibson}%
}{%
{Gibson}%
}{%
{\protect \APACyear {2018}}%
}]{%
Gibson2018}
\APACinsertmetastar {%
Gibson2018}%
\begin{APACrefauthors}%
{Gibson}, S\BPBI E.%
\end{APACrefauthors}%
\unskip\
\newblock
\APACrefYearMonthDay{2018}{}{},
\newblock
\unskip
\newblock
\APACjournalVolNumPages{Living Reviews in Solar Physics}{15}{}{7}.
\PrintBackRefs{\CurrentBib}

\bibitem [\protect \citeauthoryear {%
{Hiei}%
, {Ishiguro}%
, {Kosugi}%
\BCBL {}\ \BBA {} {Shibasaki}%
}{%
{Hiei}%
\ \protect \BOthers {.}}{%
{\protect \APACyear {1986}}%
}]{%
Hiei1986}
\APACinsertmetastar {%
Hiei1986}%
\begin{APACrefauthors}%
{Hiei}, E.%
, {Ishiguro}, M.%
, {Kosugi}, T.%
\BCBL {}\ \BBA {} {Shibasaki}, K.%
\end{APACrefauthors}%
\unskip\
\newblock
\APACrefYearMonthDay{1986}{}{},
\newblock
{\BBOQ}\APACrefatitle {{Dark filaments observed at 8.3 mm and 3.1 mm
  wavelengths.}} {{Dark filaments observed at 8.3 mm and 3.1 mm
  wavelengths.}}{\BBCQ}
\newblock
\BIn{} \APACrefbtitle {NASA Conference Publication} {NASA Conference
  Publication}\ \BVOL\ 2442, \BPGS\ 109-116, {\it Dark filaments observed at
  8.3 mm and 3.1 mm wavelengths}.
\newblock
\APACaddressPublisher{}{NASA, Scientific and Technical Information Office,
  Washington}.
\PrintBackRefs{\CurrentBib}

\bibitem [\protect \citeauthoryear {%
{Hurford}%
}{%
{Hurford}%
}{%
{\protect \APACyear {1992}}%
}]{%
Hurford1992}
\APACinsertmetastar {%
Hurford1992}%
\begin{APACrefauthors}%
{Hurford}, G.%
\end{APACrefauthors}%
\unskip\
\newblock
\APACrefYearMonthDay{1992}{}{},
\newblock
{\BBOQ}\APACrefatitle {{Solar Radio Observations}} {{Solar Radio
  Observations}}.{\BBCQ}
\newblock
\BIn{} J\BPBI T.~{Schmelz}\ \BBA {} J\BPBI C.~{Brown}\ (\BEDS), \APACrefbtitle
  {NATO Advanced Science Institutes (ASI) Series C} {NATO Advanced Science
  Institutes (ASI) Series C}\ \BVOL~373, \BPG~297.
\newblock
\APACaddressPublisher{}{Kluwer Academic Publishers, Dordrecht}.
\PrintBackRefs{\CurrentBib}

\bibitem [\protect \citeauthoryear {%
{Ivezi\'c}%
, {Connolly}%
, {VanderPlas}%
\BCBL {}\ \BBA {} {Gray}%
}{%
{Ivezi\'c}%
\ \protect \BOthers {.}}{%
{\protect \APACyear {2014}}%
}]{%
Ivezic2014}
\APACinsertmetastar {%
Ivezic2014}%
\begin{APACrefauthors}%
{Ivezi\'c}, {\v Z}.%
, {Connolly}, A\BPBI J.%
, {VanderPlas}, J\BPBI T.%
\BCBL {}\ \BBA {} {Gray}, A.%
\end{APACrefauthors}%
\unskip\
\newblock
\APACrefYear{2014},
\newblock
\APACrefbtitle {{Statistics, Data Mining, and Machine Learning in Astronomy: A
  Practical Python Guide for the Analysis of Survey Data}} {{Statistics, Data
  Mining, and Machine Learning in Astronomy: A Practical Python Guide for the
  Analysis of Survey Data}}\ (\PrintOrdinal{STU - Student}\ \BEd).
\newblock
\APACaddressPublisher{}{Princeton University Press}.
\PrintBackRefs{\CurrentBib}

\bibitem [\protect \citeauthoryear {%
{Jensen}%
\ \BBA {} {Wiik}%
}{%
{Jensen}%
\ \BBA {} {Wiik}%
}{%
{\protect \APACyear {1990}}%
}]{%
Jensen1990}
\APACinsertmetastar {%
Jensen1990}%
\begin{APACrefauthors}%
{Jensen}, E.%
\BCBT {}\ \BBA {} {Wiik}, J\BPBI E.%
\end{APACrefauthors}%
\unskip\
\newblock
\APACrefYearMonthDay{1990}{}{},
\newblock
{\BBOQ}\APACrefatitle {{Plasma Parameters in Quiescent Prominences}} {{Plasma
  Parameters in Quiescent Prominences}}.{\BBCQ}
\newblock
\BIn{} V.~{Ruzdjak}\ \BBA {} E.~{Tandberg-Hanssen}\ (\BEDS), \APACrefbtitle
  {IAU Colloq. 117: Dynamics of Quiescent Prominences} {IAU Colloq. 117:
  Dynamics of Quiescent Prominences}\ \BVOL~363, \BPG~298.
\PrintBackRefs{\CurrentBib}

\bibitem [\protect \citeauthoryear {%
{Kundu}%
, {Fuerst}%
, {Hirth}%
\BCBL {}\ \BBA {} {Butz}%
}{%
{Kundu}%
\ \protect \BOthers {.}}{%
{\protect \APACyear {1978}}%
}]{%
Kundu1978}
\APACinsertmetastar {%
Kundu1978}%
\begin{APACrefauthors}%
{Kundu}, M\BPBI R.%
, {Fuerst}, E.%
, {Hirth}, W.%
\BCBL {}\ \BBA {} {Butz}, M.%
\end{APACrefauthors}%
\unskip\
\newblock
\APACrefYearMonthDay{1978}{}{},
\newblock
\unskip
\newblock
\APACjournalVolNumPages{\aap}{62}{}{431-437}.
\PrintBackRefs{\CurrentBib}

\bibitem [\protect \citeauthoryear {%
{Lites}%
\ \protect \BOthers {.}}{%
{Lites}%
\ \protect \BOthers {.}}{%
{\protect \APACyear {2010}}%
}]{%
Lites2010}
\APACinsertmetastar {%
Lites2010}%
\begin{APACrefauthors}%
{Lites}, B\BPBI W.%
, {Kubo}, M.%
, {Berger}, T.%
\ et al.\end{APACrefauthors}%
\unskip\
\newblock
\APACrefYearMonthDay{2010}{}{},
\newblock
\unskip
\newblock
\APACjournalVolNumPages{\apj}{718}{}{474-487}.
\PrintBackRefs{\CurrentBib}

\bibitem [\protect \citeauthoryear {%
{Matkovi{\'c}}%
\ \protect \BOthers {.}}{%
{Matkovi{\'c}}%
\ \protect \BOthers {.}}{%
{\protect \APACyear {2024}}%
}]{%
Matkovic2024}
\APACinsertmetastar {%
Matkovic2024}%
\begin{APACrefauthors}%
{Matkovi{\'c}}, F.%
, {Braj{\v{s}}a}, R.%
, {Kuhar}, M.%
\ et al.\end{APACrefauthors}%
\unskip\
\newblock
\APACrefYearMonthDay{2024}{}{},
\newblock
\unskip
\newblock
\APACjournalVolNumPages{Astronomische Nachrichten}{345}{5}{e20230149}.
\PrintBackRefs{\CurrentBib}

\bibitem [\protect \citeauthoryear {%
{Okada}%
\ \protect \BOthers {.}}{%
{Okada}%
\ \protect \BOthers {.}}{%
{\protect \APACyear {2020}}%
}]{%
Okada2020}
\APACinsertmetastar {%
Okada2020}%
\begin{APACrefauthors}%
{Okada}, S.%
, {Ichimoto}, K.%
, {Machida}, A.%
, {Tokuda}, S.%
, {Huang}, Y.%
\BCBL {}\ \BBA {} {UeNo}, S.%
\end{APACrefauthors}%
\unskip\
\newblock
\APACrefYearMonthDay{2020}{}{},
\newblock
\unskip
\newblock
\APACjournalVolNumPages{Publ. Astron. Soc. Japan}{72}{}{1-17}.
\PrintBackRefs{\CurrentBib}

\bibitem [\protect \citeauthoryear {%
{Parenti}%
}{%
{Parenti}%
}{%
{\protect \APACyear {2014}}%
}]{%
Parenti2014}
\APACinsertmetastar {%
Parenti2014}%
\begin{APACrefauthors}%
{Parenti}, S.%
\end{APACrefauthors}%
\unskip\
\newblock
\APACrefYearMonthDay{2014}{}{},
\newblock
\unskip
\newblock
\APACjournalVolNumPages{Living Reviews in Solar Physics}{11}{}{1}.
\newblock
\begin{APACrefDOI} \doi{10.12942/lrsp-2014-1} \end{APACrefDOI}
\PrintBackRefs{\CurrentBib}

\bibitem [\protect \citeauthoryear {%
{Ru{\v z}djak}%
\ \BBA {} {Tandberg-Hanssen}%
}{%
{Ru{\v z}djak}%
\ \BBA {} {Tandberg-Hanssen}%
}{%
{\protect \APACyear {1990}}%
}]{%
Ruzdjak1990}
\APACinsertmetastar {%
Ruzdjak1990}%
\begin{APACrefauthors}%
{Ru{\v z}djak}, V.%
\BCBT {}\ \BBA {} {Tandberg-Hanssen}, E.%
\end{APACrefauthors}%
\unskip\
\newblock
\APACrefYear{1990},
\newblock
\APACrefbtitle {{Dynamics of Quiescent Prominences: IAU Coll.117, 1990}}
  {{Dynamics of Quiescent Prominences: IAU Coll.117, 1990}}\ (\BVOL~363).
\PrintBackRefs{\CurrentBib}

\bibitem [\protect \citeauthoryear {%
{Rybicki}%
\ \BBA {} {Lightman}%
}{%
{Rybicki}%
\ \BBA {} {Lightman}%
}{%
{\protect \APACyear {1985}}%
}]{%
Rybicki1985}
\APACinsertmetastar {%
Rybicki1985}%
\begin{APACrefauthors}%
{Rybicki}, G\BPBI B.%
\BCBT {}\ \BBA {} {Lightman}, A\BPBI P.%
\end{APACrefauthors}%
\unskip\
\newblock
\APACrefYear{1985},
\newblock
\APACrefbtitle {{Radiative Processes in Astrophysics, Wiley-VCH}} {{Radiative
  Processes in Astrophysics, Wiley-VCH}}.
\PrintBackRefs{\CurrentBib}

\bibitem [\protect \citeauthoryear {%
{Selhorst}%
, {Silva}%
\BCBL {}\ \BBA {} {Costa}%
}{%
{Selhorst}%
\ \protect \BOthers {.}}{%
{\protect \APACyear {2005}}%
}]{%
selhorst2005}
\APACinsertmetastar {%
selhorst2005}%
\begin{APACrefauthors}%
{Selhorst}, C\BPBI L.%
, {Silva}, A\BPBI V\BPBI R.%
\BCBL {}\ \BBA {} {Costa}, J\BPBI E\BPBI R.%
\end{APACrefauthors}%
\unskip\
\newblock
\APACrefYearMonthDay{2005}{}{},
\newblock
\unskip
\newblock
\APACjournalVolNumPages{\aap}{433}{}{365-374}.
\PrintBackRefs{\CurrentBib}

\bibitem [\protect \citeauthoryear {%
{Selhorst}%
\ \protect \BOthers {.}}{%
{Selhorst}%
\ \protect \BOthers {.}}{%
{\protect \APACyear {2019}}%
}]{%
Selhorst2019}
\APACinsertmetastar {%
Selhorst2019}%
\begin{APACrefauthors}%
{Selhorst}, C\BPBI L.%
, {Sim{\~o}es}, P\BPBI J\BPBI A.%
, {Braj{\v{s}}a}, R.%
\ et al.\end{APACrefauthors}%
\unskip\
\newblock
\APACrefYearMonthDay{2019}{}{},
\newblock
\unskip
\newblock
\APACjournalVolNumPages{\apj}{871}{}{45}.
\PrintBackRefs{\CurrentBib}

\bibitem [\protect \citeauthoryear {%
{Selhorst}%
\ \protect \BOthers {.}}{%
{Selhorst}%
\ \protect \BOthers {.}}{%
{\protect \APACyear {2017}}%
}]{%
selhorst2017}
\APACinsertmetastar {%
selhorst2017}%
\begin{APACrefauthors}%
{Selhorst}, C\BPBI L.%
, {Sim{\~o}es}, P\BPBI J\BPBI A.%
, {Oliveira e Silva}, A\BPBI J.%
, {Gim{\'e}nez de Castro}, C\BPBI G.%
, {Costa}, J\BPBI E\BPBI R.%
\BCBL {}\ \BBA {} {Valio}, A.%
\end{APACrefauthors}%
\unskip\
\newblock
\APACrefYearMonthDay{2017}{}{},
\newblock
\unskip
\newblock
\APACjournalVolNumPages{\apj}{851}{}{146}.
\PrintBackRefs{\CurrentBib}

\bibitem [\protect \citeauthoryear {%
{Shimojo}%
\ \protect \BOthers {.}}{%
{Shimojo}%
\ \protect \BOthers {.}}{%
{\protect \APACyear {2017}}%
}]{%
Shimojo2017}
\APACinsertmetastar {%
Shimojo2017}%
\begin{APACrefauthors}%
{Shimojo}, M.%
, {Bastian}, T.%
, {Hales}, A.%
\ et al.\end{APACrefauthors}%
\unskip\
\newblock
\APACrefYearMonthDay{2017}{}{},
\newblock
\unskip
\newblock
\APACjournalVolNumPages{\solphys}{292}{}{87}.
\PrintBackRefs{\CurrentBib}

\bibitem [\protect \citeauthoryear {%
{Sim{\~o}es}%
\ \protect \BOthers {.}}{%
{Sim{\~o}es}%
\ \protect \BOthers {.}}{%
{\protect \APACyear {2017}}%
}]{%
Simoes2017}
\APACinsertmetastar {%
Simoes2017}%
\begin{APACrefauthors}%
{Sim{\~o}es}, P\BPBI J\BPBI A.%
, {Kerr}, G\BPBI S.%
, {Fletcher}, L.%
, {Hudson}, H\BPBI S.%
, {Gim{\'e}nez de Castro}, C\BPBI G.%
\BCBL {}\ \BBA {} {Penn}, M.%
\end{APACrefauthors}%
\unskip\
\newblock
\APACrefYearMonthDay{2017}{}{},
\newblock
\unskip
\newblock
\APACjournalVolNumPages{\aap}{605}{}{A125}.
\PrintBackRefs{\CurrentBib}

\bibitem [\protect \citeauthoryear {%
{Sudar}%
, {Braj{\v s}a}%
, {Skoki\'c}%
\BCBL {}\ \BBA {} {Benz}%
}{%
{Sudar}%
\ \protect \BOthers {.}}{%
{\protect \APACyear {2019}}%
}]{%
Sudar2019}
\APACinsertmetastar {%
Sudar2019}%
\begin{APACrefauthors}%
{Sudar}, D.%
, {Braj{\v s}a}, R.%
, {Skoki\'c}, I.%
\BCBL {}\ \BBA {} {Benz}, A\BPBI O.%
\end{APACrefauthors}%
\unskip\
\newblock
\APACrefYearMonthDay{2019}{}{},
\newblock
\unskip
\newblock
\APACjournalVolNumPages{\solphys}{294}{}{163}.
\PrintBackRefs{\CurrentBib}

\bibitem [\protect \citeauthoryear {%
{Tandberg-Hanssen}%
}{%
{Tandberg-Hanssen}%
}{%
{\protect \APACyear {1995}}%
}]{%
Tandberg1995}
\APACinsertmetastar {%
Tandberg1995}%
\begin{APACrefauthors}%
{Tandberg-Hanssen}, E.%
\end{APACrefauthors}%
\unskip\
\newblock
\APACrefYear{1995},
\newblock
\APACrefbtitle {{The nature of solar prominences, Astrophysics and Space
  Science Library, Dordrecht: Kluwer Academic Publishers}} {{The nature of
  solar prominences, Astrophysics and Space Science Library, Dordrecht: Kluwer
  Academic Publishers}}\ (\BVOL~199).
\PrintBackRefs{\CurrentBib}

\bibitem [\protect \citeauthoryear {%
{Urpo}%
, {Pohjolainen}%
, {Heikkil{\"a}}%
\BCBL {}\ \BBA {} {Wiik}%
}{%
{Urpo}%
\ \protect \BOthers {.}}{%
{\protect \APACyear {1997}}%
}]{%
Urpo1997}
\APACinsertmetastar {%
Urpo1997}%
\begin{APACrefauthors}%
{Urpo}, S.%
, {Pohjolainen}, S.%
, {Heikkil{\"a}}, J.%
\BCBL {}\ \BBA {} {Wiik}, K.%
\end{APACrefauthors}%
\unskip\
\newblock
\APACrefYear{1997},
\newblock
\APACrefbtitle {{Solar Observations at Mets{\"a}hovi in 1994-1995, Hels. Univ.
  Technol., Mets{\"a}hovi Radio Res. Stn., Rep. Ser. A, No. 26}} {{Solar
  Observations at Mets{\"a}hovi in 1994-1995, Hels. Univ. Technol.,
  Mets{\"a}hovi Radio Res. Stn., Rep. Ser. A, No. 26}}.
\PrintBackRefs{\CurrentBib}

\bibitem [\protect \citeauthoryear {%
{van Hoof}%
\ \protect \BOthers {.}}{%
{van Hoof}%
\ \protect \BOthers {.}}{%
{\protect \APACyear {2014}}%
}]{%
vanHoof2014}
\APACinsertmetastar {%
vanHoof2014}%
\begin{APACrefauthors}%
{van Hoof}, P\BPBI A\BPBI M.%
, {Williams}, R\BPBI J\BPBI R.%
, {Volk}, K.%
\ et al.\end{APACrefauthors}%
\unskip\
\newblock
\APACrefYearMonthDay{2014}{}{},
\newblock
\unskip
\newblock
\APACjournalVolNumPages{\mnras}{444}{}{420-428}.
\PrintBackRefs{\CurrentBib}

\bibitem [\protect \citeauthoryear {%
{Vernazza}%
, {Avrett}%
\BCBL {}\ \BBA {} {Loeser}%
}{%
{Vernazza}%
\ \protect \BOthers {.}}{%
{\protect \APACyear {1981}}%
}]{%
Vernazza1981}
\APACinsertmetastar {%
Vernazza1981}%
\begin{APACrefauthors}%
{Vernazza}, J\BPBI E.%
, {Avrett}, E\BPBI H.%
\BCBL {}\ \BBA {} {Loeser}, R.%
\end{APACrefauthors}%
\unskip\
\newblock
\APACrefYearMonthDay{1981}{}{},
\newblock
\unskip
\newblock
\APACjournalVolNumPages{\apjs}{45}{}{635-725}.
\PrintBackRefs{\CurrentBib}

\bibitem [\protect \citeauthoryear {%
{Vr{\v s}nak}%
\ \protect \BOthers {.}}{%
{Vr{\v s}nak}%
\ \protect \BOthers {.}}{%
{\protect \APACyear {1992}}%
}]{%
Vrsnak1992}
\APACinsertmetastar {%
Vrsnak1992}%
\begin{APACrefauthors}%
{Vr{\v s}nak}, B.%
, {Pohjolainen}, S.%
, {Urpo}, S.%
\ et al.\end{APACrefauthors}%
\unskip\
\newblock
\APACrefYearMonthDay{1992}{}{},
\newblock
\unskip
\newblock
\APACjournalVolNumPages{\solphys}{137}{}{67-86}.
\PrintBackRefs{\CurrentBib}

\bibitem [\protect \citeauthoryear {%
{Wedemeyer}%
\ \protect \BOthers {.}}{%
{Wedemeyer}%
\ \protect \BOthers {.}}{%
{\protect \APACyear {2024}}%
}]{%
Wedemeyer2024}
\APACinsertmetastar {%
Wedemeyer2024}%
\begin{APACrefauthors}%
{Wedemeyer}, S.%
, {Barta}, M.%
, {Braj{\v{s}}a}, R.%
\ et al.\end{APACrefauthors}%
\unskip\
\newblock
\APACrefYearMonthDay{2024}{}{},
\newblock
\unskip
\newblock
\APACjournalVolNumPages{Open Research Europe}{4}{}{140}.
\PrintBackRefs{\CurrentBib}

\bibitem [\protect \citeauthoryear {%
{Wedemeyer}%
\ \protect \BOthers {.}}{%
{Wedemeyer}%
\ \protect \BOthers {.}}{%
{\protect \APACyear {2016}}%
}]{%
Wedemeyer2016}
\APACinsertmetastar {%
Wedemeyer2016}%
\begin{APACrefauthors}%
{Wedemeyer}, S.%
, {Bastian}, T.%
, {Braj\v sa}, R.%
\ et al.\end{APACrefauthors}%
\unskip\
\newblock
\APACrefYearMonthDay{2016}{}{},
\newblock
\unskip
\newblock
\APACjournalVolNumPages{Space Science Reviews}{198}{}{1}.
\PrintBackRefs{\CurrentBib}

\bibitem [\protect \citeauthoryear {%
{Weinberg}%
}{%
{Weinberg}%
}{%
{\protect \APACyear {2020}}%
}]{%
Weinberg2020}
\APACinsertmetastar {%
Weinberg2020}%
\begin{APACrefauthors}%
{Weinberg}, S.%
\end{APACrefauthors}%
\unskip\
\newblock
\APACrefYear{2020},
\newblock
\APACrefbtitle {{Lectures on Astrophysics, Cambridge University Press,
  Cambridge}} {{Lectures on Astrophysics, Cambridge University Press,
  Cambridge}}.
\PrintBackRefs{\CurrentBib}

\bibitem [\protect \citeauthoryear {%
{White}%
}{%
{White}%
}{%
{\protect \APACyear {2002}}%
}]{%
White2002}
\APACinsertmetastar {%
White2002}%
\begin{APACrefauthors}%
{White}, S\BPBI M.%
\end{APACrefauthors}%
\unskip\
\newblock
\APACrefYearMonthDay{2002}{}{},
\newblock
{\BBOQ}\APACrefatitle {{The Solar Atmosphere at Radio Wavelengths}} {{The Solar
  Atmosphere at Radio Wavelengths}}.{\BBCQ}
\newblock
\BIn{} F.~{Favata}\ \BBA {} J\BPBI J.~{Drake}\ (\BEDS), \APACrefbtitle {Stellar
  Coronae in the Chandra and XMM-NEWTON Era} {Stellar Coronae in the Chandra
  and XMM-NEWTON Era}\ \BVOL~277, \BPG~299.
\newblock
\APACaddressPublisher{}{ASP, San Francisco}.
\PrintBackRefs{\CurrentBib}

\bibitem [\protect \citeauthoryear {%
{White}%
\ \protect \BOthers {.}}{%
{White}%
\ \protect \BOthers {.}}{%
{\protect \APACyear {2017}}%
}]{%
White2017}
\APACinsertmetastar {%
White2017}%
\begin{APACrefauthors}%
{White}, S\BPBI M.%
, {Iwai}, K.%
, {Phillips}, N.%
\ et al.\end{APACrefauthors}%
\unskip\
\newblock
\APACrefYearMonthDay{2017}{}{},
\newblock
\unskip
\newblock
\APACjournalVolNumPages{\solphys}{292}{}{88}.
\PrintBackRefs{\CurrentBib}

\bibitem [\protect \citeauthoryear {%
{Wilson}%
, {Rohlfs}%
\BCBL {}\ \BBA {} {H\"uttemeister}%
}{%
{Wilson}%
\ \protect \BOthers {.}}{%
{\protect \APACyear {2013}}%
}]{%
Wilson2013}
\APACinsertmetastar {%
Wilson2013}%
\begin{APACrefauthors}%
{Wilson}, T\BPBI L.%
, {Rohlfs}, K.%
\BCBL {}\ \BBA {} {H\"uttemeister}, S.%
\end{APACrefauthors}%
\unskip\
\newblock
\APACrefYear{2013},
\newblock
\APACrefbtitle {{Tools of Radio Astronomy}} {{Tools of Radio Astronomy}}\
  (\PrintOrdinal{6}\ \BEd).
\newblock
\APACaddressPublisher{}{Springer Berlin Heidelberg}.
\PrintBackRefs{\CurrentBib}

\bibitem [\protect \citeauthoryear {%
{Yang}%
\ \protect \BOthers {.}}{%
{Yang}%
\ \protect \BOthers {.}}{%
{\protect \APACyear {2024}}%
}]{%
Yang2024}
\APACinsertmetastar {%
Yang2024}%
\begin{APACrefauthors}%
{Yang}, Z.%
, {Tian}, H.%
, {Tomczyk}, S.%
, {Liu}, X.%
, {Gibson}, S.%
, {Morton}, R\BPBI J.%
\BCBL {}\ \BBA {} {Downs}, C.%
\end{APACrefauthors}%
\unskip\
\newblock
\APACrefYearMonthDay{2024}{}{},
\newblock
\unskip
\newblock
\APACjournalVolNumPages{Science}{386}{6717}{76-82}.
\PrintBackRefs{\CurrentBib}

\bibitem [\protect \citeauthoryear {%
{Zirin}%
}{%
{Zirin}%
}{%
{\protect \APACyear {1988}}%
}]{%
Zirin1988}
\APACinsertmetastar {%
Zirin1988}%
\begin{APACrefauthors}%
{Zirin}, H.%
\end{APACrefauthors}%
\unskip\
\newblock
\APACrefYear{1988},
\newblock
\APACrefbtitle {{Astrophysics of the Sun, Cambridge University Press,
  Cambridge}} {{Astrophysics of the Sun, Cambridge University Press,
  Cambridge}}.
\PrintBackRefs{\CurrentBib}

\end{thebibliography}

\end{document}